\documentclass[amsmath,amssymb,superscriptaddress,epsfig,aps,pra,twocolumn]{revtex4-1}
\usepackage{amsmath}
\usepackage{epsfig}
\usepackage{graphicx}
\usepackage{xcolor}
\usepackage{amssymb}
\usepackage{epstopdf}
\usepackage{array}
\usepackage{tabularx}
\usepackage[utf8]{inputenc}
\usepackage[T1]{fontenc}
\usepackage[colorlinks=True, linkcolor=blue, citecolor=blue]{hyperref}
\begin{document}
\title{Rayleigh-Taylor, Kelvin-Helmholtz and immiscible to miscible quenching
 instabilities in binary Bose-Einstein condensates}
\author{R. Kishor Kumar}
\affiliation{Department of Physics, Centre for Quantum Science and Dodd-Walls 
Centre for Photonic and Quantum Technologies, University of Otago, Dunedin 9054, 
New Zealand.}
\author{S. Sabari}
\affiliation{Instituto de F\'isica Te\'orica, Universidade Estadual Paulista 
(UNESP), 01140-070 S\~ao Paulo, SP, Brazil}
\author{Arnaldo Gammal}
\affiliation{Instituto de F\'{i}sica, Universidade de S\~{a}o Paulo,
05508-090 S\~{a}o Paulo, SP, Brazil}
\author{Lauro Tomio}
\affiliation{Instituto de F\'isica Te\'orica, Universidade Estadual Paulista 
(UNESP), 01140-070 S\~ao Paulo, SP, Brazil}
\affiliation{International Physics Center, Institute of Physics, University of 
Brasilia, 70910-900 Brasília, DF, Brazil}
\date{\today}
\begin{abstract}
We investigate three kinds of instabilities in binary immiscible homogeneous Bose-Einstein 
condensate, considering rubidium isotopes $^{85}$Rb and $^{87}$Rb confined in two-dimensional 
circular box. 
Rayleigh-Taylor (RT) and Kelvin-Helmholtz (KH) instability types are studied under strong 
perturbations. Without external perturbation, instabilities 
are also probed by immiscible to miscible quenching transition (IMQT),
under two different initial configurations. Our numerical simulations show that 
all such instability dynamics are dominated by large vortex production and 
sound-wave (phonon) propagation.
For long-term propagation, vortex dynamics become dominant
over sound waves in the KH instability, while sound-wave excitations predominate 
in the other cases.
For all the dynamical simulations, the emergence of possible 
scaling laws are investigated for the compressible and incompressible parts of the 
kinetic energy spectra, in terms of the wavenumber $k$. The corresponding 
results are compared with the classical Kolmogorov scalings, $k^{-5/3}$ and 
$k^{-3}$, for turbulence, which are observed in the kinetic energy spectra 
at some specific time intervals. 
\end{abstract}
\flushbottom
\maketitle
\section{Introduction}
\label{secI}
In classical fluid dynamics, two regimes can be characterized by the flow of a fluid, 
being laminar or turbulent. The first occurs when viscous forces dominate, leading to 
smooth and constant motion, with the turbulent flow dominated by inertial forces, 
creating vortices, chaotic eddies, and instabilities. The Reynolds number serves as 
a guide to measuring the effect of fluid friction and viscosity in the fluid, being 
low for constant smooth fluid motion (laminar) and high for the case where the flow is 
turbulent~\cite{1883-Reynolds,1908-Sommerfeld,2010-Eckert}. Most flows observed in 
nature and physical systems are turbulent. 
The structure of turbulence in classical incompressible fluids was originally proposed by 
Kolmogorov~\cite{Kolmogorov1941} in 1941 as related to large Reynolds numbers. 
This and other studies by Kolmogorov on turbulence are detailed in a review by 
Frisch~\cite{1995Frisch}, together with the related investigations by several other authors.
In quantum fluids, such as superfluid helium and Bose-Einstein condensate (BEC) atomic systems, 
turbulent motions have emerged as a new interdisciplinary research topic, named quantum 
turbulence (QT) in the literature~\cite{1986Donnelly}. For an ideal zero-temperature superfluid, 
the problem in the characterization of turbulence by the Reynolds number~\cite{2008Barenghi} 
was discussed in Refs.~\cite{2012Bradley,2015Reeves}, in which the authors define a superfluid 
Reynolds number via dynamical similarity, to identify a regime transition to turbulence.
Also, by considering heat transport and thermal waves in laminar and turbulent superfluid 
helium~\cite{1991Donnelly}, a quantum analogous Reynolds number for the transition 
to superfluid turbulence was defined in Ref.~\cite{2013Jou}, with a related discussion 
in the context of BEC formalism in Ref.~\cite{2018Mongiovi}. 
Currently, the studies on QT can be followed by several available
works and reviews~\cite{2002Vinen,2011Paoletti,2013Tsubota,2013Nemirovskii,
2014Kwon, 2016Barenghi,2016Tsatsos},
recently updated in Ref.~\cite{2023Barenghi}. The quite recent works reported in 
Refs.~\cite{2025Zhao,2025Fischer,2025Simjanovski} are also indicative of the 
actual interest in the analyses of QT in BEC. 
Some related numerical analyses and benchmark 
high-performing computer simulations are provided in Refs.~\cite{2017Tsubota,2021Kobayashi}.
Turbulence is associated with stochastic movements of vortices in a fluid, whose properties 
differ in classical and quantum physics. Therefore, it is rooted in the fundamental 
differences in how vorticity behaves in these two regimes. In the quantum regime, vortices 
have quantized circulation, and the flow has negligible viscosity. 
Vortices do not decay by viscous diffusion but through other mechanisms like reconnections 
and phonon emission, as shown in an
experimental and numerical study of 3D quantum vortex interactions 
in a cigar-shaped atomic BEC~\cite{2017Serafini}. 
In the classical regime, vortices can have any circulation value, 
and vorticity decays through viscous diffusion, leading to the dissipation of turbulent energy.
These differences make quantum turbulence a fascinating area of study, with 
implications for understanding superfluidity, quantum fluids, and the behavior of 
matter at very low temperatures.

The first experimental observation of QT was reported by considering the $^4$He 
superfluid~\cite{Vinen1957}, with more recent experiments being extended to atomic 
BECs~\cite{Bagnato2009,2014White,Navon2016,Navon2019,2024Karailiev,2022Reeves}. 
The BEC experiments and related studies have received significant attention due to the 
advanced techniques available in cold-atom laboratories, which allow for precise control 
over the condensate parameters. For instance, in BEC researchers can manipulate the 
trapping potential, interaction strength, and temperature with high accuracy, enabling 
detailed studies of quantum turbulence.
The characterization of a turbulent fluid occurs through spectral analysis,
with the energy being distributed across different length scales, whereas in a non-turbulent
fluid, there is no significant energy transfer across scales.
The evolution of large clusters with $^{87}$Rb BECs has been demonstrated in experiments 
reported in Ref.~\cite{Gauthier2019}, with large-scale flow from turbulence being studied 
experimentally by considering a two-dimensional (2D) superfluid~\cite{Johnstone2019}.
Identified in the energy spectrum of these experiments, 
the emergence of the classical Kolmogorov's scaling~\cite{Kolmogorov1941} was also previously 
observed in studies on turbulence at low-temperature superfluid flows~\cite{1997Nore}.
These experiments collectively advance our understanding of quantum turbulence, bridging 
the gap between classical and quantum fluid dynamics. The observation of Kolmogorov's 
scaling in these systems is particularly noteworthy, as it suggests that some aspects 
of turbulence are universal, transcending the classical and quantum divide.

Further, theoretical studies on QT have been established by using appropriate versions 
of the mean-field Gross-Pitaevskii (GP) formalism, as one can follow from 
Refs.~\cite{1997Nore,2005Kobayashi,2009Yepez,2011Baggaley}.
In this regard, one should note that, in the kinetic energy spectrum 
over the wavenumber $k$ of quantum fluids, the occurrence of the inverse energy cascade 
phenomenon was clearly verified in Refs.~\cite{2013Reeves,2014Reeves}, 
through the analyses of QT in forced 2D, and related signatures of coherent emerging 
vortex structures. Consistent with Kolmogorov's scaling, the vortex dynamics follow a 
$k^{-5/3}$ power law in the infrared region of the spectrum and $k^{-3}$ in the 
ultraviolet region.  Such scaling laws provide insights into the nature of 
energy transfer in QT, with large-scale vortices dominating the dynamics in the 
infrared region, while the $k^{-3}$ scaling in the ultraviolet region indicates 
a turbulent cascade similar to that observed in classical turbulence, where energy 
is transferred from larger to smaller scales.

Meanwhile, the ongoing experimental studies with binary atomic mixtures and  
hyperfine spin states of the same atom~\cite{1998Hall} provide motivations to 
extend such studies to turbulence and vortex patterns in BEC multi-component 
mixtures. They are of great interest due to the miscibility 
properties~\cite{2002Pethick,2017kumar}. In particular, the phase-separated binary 
mixtures show a rich variety of pattern formations, recognized as similar to the 
classical Rayleigh-Taylor (RT)~\cite{1900Rayleigh,1950Taylor} and Kelvin-Helmholtz 
(KH)~\cite{1871Thomson,1868Helmholtz} instabilities. The RT instability is overviewed
in Ref.~\cite{1984sharp} and more recently in Ref.~\cite{2020Banerjee}; 
whereas the KH was first studied by considering instability in 
superfluids~\cite{2002Blaauwgeers,2006Finne}. 
When considering BEC mixtures, the studies on these instabilities can be followed by 
plenty of works from the last two decades up to now, exemplified by 
Refs.~\cite{2006Berloff,Sasaki2009,Takeuchi2010,2010Gautam,Bezett2010,Kobyakov2011,Kobyakov2014,
2016Fujimoto, 2021Mithun, 2021Kokubo, 2023Saboo, 2023Silva, 2024Kadokura}.
From classical fluid experiments, RT instabilities are known to start at the interface 
between two plane-parallel immiscible fluids under the gravity field, with the denser 
fluid layer at the top of the less dense one. As the equilibrium is broken, the fluid 
at the top moves downward, with an equal volume of the lighter one pushing upward, 
resulting in mushroom head formations of the denser fluid inside the space first occupied by
the less dense one. Distinguishable from RT, the KH instabilities occur when there is a 
difference in speed between the two fluids at the interface. These instabilities are 
investigated theoretically by using mean field and Bogoliubov theories. Analytical tools 
are developed for classical studies on hydrodynamic instabilities and turbulence, which 
can be followed by the overview provided in Ref.~\cite{2024Zhou}.
However, no extensive studies discuss the 
QT for homogeneous binary immiscible mixtures from the perspective of analyzing Kolmogorov's 
spectrum, which could reveal some relation between QT and classical turbulence. 
In conjunction with Kolmogorov turbulence, a recent study was also reported in  
Ref.~\cite{2024Kadokura}, by considering a strongly stirred immiscible mixture of 
two-components, in which the authors extract power-law behaviors associated with the highly 
or slightly immiscible conditions.

Given current cold-atom experimental activities on dipolar systems 
(see \cite{2024Bigagli}, on the observation of dipolar molecules, and references therein),
some of us have also considered QT in dipolar BECs, generated by an external penetrable 
Gaussian-type circularly moving obstacle that produces vortex-antivortex pairs~\cite{Sabari2024}, 
following related studies with linearly-moving obstacles~\cite{2018Sabari}. 
Along these lines, binary BEC mixtures have also been studied 
with the assumptions of mass-imbalanced components confined in quasi-two-dimensional (quasi-2D)
pancake-like trap potential slightly perturbed elliptically by a time-dependent periodic 
potential~\cite{2023Silva,Lauro2024}. In the analysis of turbulence in mixtures of quantum 
fluids when considering quantized vortices~\cite{1991Donnelly}, another aspect of interest 
is to identify the complexity by the corresponding fractal dimensions and scale 
distributions~\cite{2001Kivotides,2020Koniakhin}, which can also be done by following 
some classical investigations~\cite{1985Constantin,2009Rakhshandehroo}.

In this work, we aim to perform numerical investigations on the emergence of 
instabilities and quantum turbulence in binary BEC mixtures confined in a 
quasi-2D circular box, considering different possible 
initial conditions for the dynamical evolution of the mixture. In all these 
cases, the coupled system is initially prepared in an immiscible stable
configuration, by solving the GP equation in the imaginary time.
Then, before following the real-time dynamical evolution, for the onset of instabilities, 
the initial non-equilibrium state is prepared by 
changing the linear and/or nonlinear interactions, as detailed in our numerical simulations.
Particular aspects of the dynamics of each kind of instability that we are reporting may 
demand further deep-focused investigation beyond our present work's scope.
Within our numerical simulations, we have assumed the coupled $^{85}$Rb-$^{87}$Rb 
system. However, the corresponding results can be easily extended to other coupled 
binary mixtures, as well as to coupled spinor states of the same atomic species, 
such as the recently reported experiments on RT 
instability considering hyperfine states of $^{23}$Na~\cite{2024Geng}. 
By assuming initial immiscible conditions, with interspecies interactions 
larger than the intraspecies ones, homogeneous density distributions will be 
considered for both spatially trapped separated components. 

In our following studies of RT and KH kinds of instabilities, the same initial 
ground-state configurations are assumed, with the mixtures kept in immiscible 
regimes along the time evolution. The RT instability is triggered by introducing 
an initial time-independent sinusoidal perturbation in the ground-state solution 
(previously prepared in imaginary time), along the $x-$direction, in the interface 
between the components initially represented by $y=0$. The perturbation in the real-time 
evolution is applied to one of the species over a short time interval. 
The simulation follows with the sinusoidal perturbation replaced by  
linearly growing potential along the $y-$direction, providing space constant forces
applied to both components in opposite directions, breaking the stability of the 
interface between the immiscible fluids. For the KH instability, 
to simulate the difference in speed between the two classical fluids at the interface, the 
additional external potentials applied to both condensed species provide constant 
spacial forces along opposite $x-$directions, parallel to the interface between 
the immiscible mixture.

Next, by extending our binary instability analyses, we try to distinguish 
the above two cases, in which external linear forces are responsible for the dynamics, 
from cases in which the dynamical instabilities are due to sudden variations of the 
nonlinear interactions. Therefore, in this third approach, we investigate the dynamical 
response of the system under an immiscible to a miscible quenching transition (IMQT), 
implemented by a sudden reduction in the ratio between the inter- and intra-species 
scattering lengths near the transition threshold. For that, the dynamics is 
explored by considering two different initial conditions for the space configuration 
and quenching transition. The interest in this case follows previous studies on phase 
separation and modulation instabilities with two-component atomic 
systems~\cite{2024Bayocboc,
Mukherjee2020,2016Eto,2004Kasamatsu,2019Thiruvalluvar}.
As will be shown, all the above-prescribed instabilities 
induce numerous vortex dipoles and turbulent flow in the condensates. 
To analyze and understand them, we calculate the corresponding compressible 
and incompressible kinetic energy parts of the spectrum, by following an 
approach detailed in Ref.~\cite{2022Bradley}. 
As is known, in a fluid mixture, the compressible part is associated with 
density fluctuations and production of sound waves (phonons), whereas the incompressible part 
is associated with vortex dynamics and rotational motion. Their analyses 
can provide information on possible universal scaling laws, which could bring some
consistency with the classical Kolmogorov's scaling for turbulence, as the studies 
provided in Ref.~\cite{2013Reeves}.

The present study is concerned with the expectation of improving our understanding 
of possible similarities and characteristic differences when comparing instabilities 
of classical and quantum fluids. In particular, significant differences are expected 
to arise due to quantization effects, which do not occur in classical fluids,
such as vortex quantization. On the other hand, the knowledge of the dynamics of 
quantum turbulence may also help us improve our understanding of  
classical fluid turbulence (which was referred to by Richard Feynman as the most
important unsolved problem in classical physics, given the lack of a 
description from first principles~\cite{1955Feynman}). Within this purpose, 
we follow similar techniques of analysis by looking for the corresponding 
Kolmogorov's scaling in the spectrum. For that, the behavior of the compressible 
and incompressible parts of the fluid is verified from large to small scales, 
in time intervals when the onset of instabilities can occur, 
such that a comparison can be established with the 
classical counterpart behavior. For longer times, we are not expecting that 
a Kolmogorov-like spectral analysis could be helpful in our comparison, 
considering that the QT dynamics should differ from classical dynamics 
due to the presence of quantized vortices and the absence of viscosity.

In the next Sect.~\ref{sec2}, the coupled GP mean-field model formalism is being 
described together with the corresponding expressions for the kinetic energy 
spectrum decomposition. The instability simulations are described and discussed in 
Sect.~\ref{sec3}, by considering the RT and KH instabilities generated by external
forces, together with IMQT instabilities obtained by 
sudden changes in the miscibility of the coupled system.
Finally, a summary with main conclusions is given in Sect.~\ref{sec4}.

\section{Mean-field model for binary BEC confined in a uniform circular box}
\label{sec2}
In our approach for the coupled BEC system, we are assuming a quasi-2D pancake-like configuration, 
with two atomic species (identified by $i=1,2$, with masses $m_i$) having the same number of atoms 
$N_i\equiv N$, strongly confined by three-dimensional (3D) harmonic traps with aspect ratios 
$\lambda_i=\omega_{i,z}/\omega_{i,\perp}$, where $\omega_{i,z}$ and $\omega_{i,\perp} 
(\equiv \omega_{i,x}=\omega_{i,y})$ are, 
respectively, the longitudinal and transverse trap frequencies for the species $i$. 
Given our units in terms of the lighter particle 1, in order to have both particles 
trapped with about the same aspect ratio $\lambda$, one needs 
$m_1\omega^2_{1z}=m_2\omega^2_{2z}$, implying $\omega_{2z}/\omega_{1z}\approx 0.99$. 
So, with trap frequencies $\omega_{i,\perp}$ 
being the same, given by $\omega_\perp\equiv \omega_{i,\perp}=2\pi\times 10\,$Hz, 
we can assume $\lambda\approx \lambda_i=50$, with $\omega_{i,z}=2\pi\times 500\,$Hz.
Within such constraints, with the binary system strongly confined to a pancake-like 
2D shape,  the original 3D formalism can be reduced to a quasi-2D one, 
by solving analytically the part of the Hamiltonian corresponding to the $z-$variable,
in which the trap is given by $V_i(z)=m_i\omega_{i,z}^2z^2/2$, implying a
constant factor in the 2D formalism. As the 2D system is strongly confined,
for practical purposes, we modify the 2D-harmonic trap by further
assuming the coupled system is confined in a uniform circular box,
as well as eventually modified according to the different model approaches under
analysis.  Relying on the experimental possibilities for tuning the two-species contact 
interactions $a_{ij}$ through Feshbach resonance mechanisms~\cite{1999Timmermans,2010Chin}, 
in all the simulations, we are assuming the contact interactions $a_{ij}$ are sufficiently 
repulsive, with identical and fixed intraspecies scattering lengths for both species,
$a_{11}=a_{22}=100a_0$ ($a_0$ is the Bohr radius), varying 
the interspecies $a_{12}$ to control the miscibility. 
Looking for sufficiently controllable instabilities, our numerical choices of 
inter- and intra-species parameters are not too far apart, with their ratio being 
close to the miscibility threshold.
As already pointed out, in the real-time evolution, the onset of the instabilities is provoked 
by different kinds of changes in the linear or nonlinear interactions, as will be detailed.
While linear perturbations trigger RT and KH instabilities, those due to IMQT are 
caused by sudden changes in nonlinear interactions, considering two types of initial 
configurations.

Following the above discussion, we have the corresponding mean-field 2D coupled GP formalism.
For convenience and computational purposes, this 2D formalism is cast in a 
dimensionless format, 
using the original harmonic trap parameters, with energy, time, and length units given, 
respectively, by $\hbar \omega_{\perp}$, $\omega_{\perp}^{-1}$, and 
$l_{\perp}\equiv \sqrt{\hbar/(m_1\omega_{\perp})}$, 
with the first species being used as the reference for the length unit. 
Correspondingly, the space and time variables are such that  
${\bf r}\to l_\perp {\bf r}$ and $t\to t/\omega_{\perp}$, when 
going to dimensionless quantities.
For details on how the 2D dimensionless formalism is reached for a mass-different
binary system, one can analogously follow the related expressions given in
Ref.~\cite{2019Kumar}.
Therefore, the coupled 2D GP equation, for the wave-function components 
$\psi_i\equiv\psi_i(x,y;t)$, 
normalized to one, $\int_{-\infty}^{\infty}dx\,dy\,|\psi _{i}|^{2}=1$, is given by
\begin{eqnarray}
\mathrm{i}\frac{\partial \psi_{i}  }{\partial t }
&=&{\bigg\{}\frac{-m_{1}}{2m_{i}}\nabla_2^2
+V_i(x,y) +\sum_{j=1,2}g_{ij}|\psi_{j} |^{2}{\bigg\}}
\psi_{i}  
\label{2d-2c}, 
\end{eqnarray}
where $\nabla_2^2\equiv \partial^2/\partial x^2 + \partial^2/\partial y^2$.
The nonlinear strengths $g_{ij}$ refer to the contact interactions, related to the two-body 
scattering lengths $a_{ij}=a_{ji}$, given by 
\begin{eqnarray}
g_{ij}\equiv \sqrt{2\pi\lambda}
\frac{m_1 a_{ij} N_j}{m_{ij}l_\perp},
\label{par}
\end{eqnarray}
where $m_{ij} \equiv m_im_j/(m_i+m_j)$ is the reduced mass. 
In Eq.~(\ref{2d-2c}), $V_i(x,y)$ is the 2D confining potential, 
initially assumed to have an identical form for both species $i=1,2$,
which will be altered by linear perturbations along 
our numerical simulations, as will be detailed below.
So, initially, we consider $V_i(x,y)$ as given by a 
uniform circular box with fixed radius $R$ and 
height $V_0$, given by
\begin{eqnarray}
V_{i}(x,y) 
&=&\left\{ \begin{array}{l}
V_0,\;{\rm for}\;\;\sqrt{x^2+y^2}>R,\\
0,\;\;{\rm for}\;\;\sqrt{x^2+y^2}\le R,
\end{array}\right.
\label{2Dtrap}
\end{eqnarray}
where $V_0$ will be considered much larger than the 
dimensionless chemical potentials, $V_0 \gg \mu_i$.
(With our units defined in terms of the transversal 
confining harmonic potential, as in Ref.~\cite{Kishor2020},
the consistency with approaches using the longitudinal frequency $\omega_z$, 
as Ref.~\cite{2012Bradley}, 
requires considering $\lambda=50$, with 
$l_\perp=\sqrt{\lambda}\,l_z$). 

Next, within our specific numerical simulations, to be 
described in the next sections, $V_i(x,y)$ will 
be altered by linear perturbations that will not have
identical form for both species $i=1,2$.
Along this work, together with the same number of atoms for both species, 
we also assume the length unit adjusted to  
$l_\perp = 3.4\,\mu$m$\,\approx 6.425 \times 10^4 a_0$, 
where $a_0$ is the Bohr radius, such that $a_{ij}$ can be conveniently given 
in terms of $a_0$.
The other fixed numerical factors are the size of the 2D circular 
box potential, defined in (\ref{2Dtrap}), which we assume $R=35$ (in units of 
$l_\perp$), the number of atoms of both components $N=N_i=2\times10^6$ and the 
equal intra species interaction $a_{ii}=100a_0$. 

\subsection{Two-component miscibility}
The condition to enter the immiscible regime corresponds to the one to minimize the
energy~\cite{2002Pethick}, given by $g_{12}^2 > g_{11} g_{22}$, for $N_1=N_2$
and $a_{11}=a_{22}>0$ [using Eq.~(\ref{par})].  
This defines the threshold parameter $\delta$, with the immiscible
necessary condition:
\begin{equation}
\delta\equiv \sqrt{\frac{g_{12}^2}{g_{11}g_{22}}}= 
\frac{m_1+m_2}{2\sqrt{m_1m_2}}\left(\frac{a_{12}}{a_{11}}\right) > 1.
\label{delta}\end{equation}
The expression at the right provides the mass-dependent critical value in terms 
of the ratio between inter- and intraspecies two-body scattering lengths for the 
miscible-immiscible transition of a homogeneous mixture. 
In the present case, this critical value is close to the equal mass case 
($a_{12}\simeq 0.99993 a_{11}$). A relation for the density overlap of the coupled 
Bose mixtures can provide a closer definition of the miscibility,
as the ones considered in Refs.~\cite{2017kumar} and \cite{Mukherjee2020}.
In the present investigation, for the density overlap, we assume the same definition as 
the one considered in Ref.~\cite{Mukherjee2020}, given by
\begin{eqnarray}
\Lambda = \frac{\left[\int \vert \psi_1 \vert^2 \vert \psi_2 \vert^2 \ dx dy \right]^2}{ \left(
\int {\vert \psi_1 \vert^4} \ dxdy \right) \left( \int { \vert \psi_2 \vert^4} \ dxdy \right) } 
,\label{eta_new}\end{eqnarray}
such that, $\Lambda=1$, for the complete density overlap; and zero for the complete immiscible case.

\subsection{Kinetic energy spectrum decomposition}
From the Eq.~(\ref{2d-2c}), the corresponding total energy for the binary system is given by
{\small \begin{eqnarray}
E(\psi_1,\psi_2)&=&\int dx dy
\sum_{i}\left\{\frac{m_1|\nabla_2\psi_i|^2}{2m_i} + V_i(x,y)|\psi_i|^2
\right\}\nonumber\\
&+&\int dx dy \sum_{i,j}\frac{g_{ij}}{2}|\psi_{j} |^{2}|\psi_{i} |^{2}
\label{GP-En}.
\end{eqnarray}
}To analyze the turbulent behavior that can occur in the coupled system, it is appropriate to 
study the corresponding decomposition of the kinetic energy spectrum for the present 2D
formalism, as detailed in Refs.~\cite{2012Bradley,2022Bradley}.
For that, in the fluid dynamics interpretation of the GP equation, we apply the 
Madelung transformation,
such that the coupled condensate wave function is given by 
$\psi_i\equiv\sqrt{n_i}\exp{({\rm i}\theta_i)}$, where $n_i\equiv n_i(x,y;t)$ 
is the density of the species $i$, with $\theta_i\equiv\theta_i(x,y;t)$ the corresponding fluid 
macroscopic phase. 
With the fluid velocity for each component $i$ being defined as 
${\bf v}_i(x,y;t)=\nabla_2\theta_i$, 
and with correspondingly density-weighted velocity given by ${\bf u}_i \equiv 
{\bf u}_i (x,y;t)\equiv \sqrt{n_i} {\bf v}_i(x,y;t)$,  
the kinetic energies of each component $i$ of the mixture can be expressed by 
\begin{eqnarray}\label{energies}
 K_{i} &=& \frac{m_1}{2m_i} \int dxdy\, |\mathrm {\bf u}_i|^2.
\end{eqnarray}

The kinetic energy is further decomposed in compressible and incompressible parts, with 
the incompressible energy primarily due to the presence of quantized vortices, and the 
compressible energy associated to density fluctuation and production of
sound waves (phonons).
We write this decomposition as 
  ${\bf u}_i=\, {\bf u}_{i,I}+ {\bf u}_{i,C}$, 
 in which the incompressible field ${\bf u}_{i,I}$ 
 satisfies $\nabla\cdot{\bf u}_{i,I} = 0$, with the  
 compressible field ${\bf u}_{i,C}$ satisfying $\nabla \times{\bf u}_{i,C} = 0$.  Therefore,
 the kinetic energy terms are decomposed as $K_{i}=K_{i,I}+K_{i,C}$, where the 
 respective compressible and incompressible parts are defined as
\begin{equation}
\left(
\begin{array}{c}
K_{i,I}\\
K_{i,C}
\end{array}\right)
= \frac{m_1}{2m_i} \int dxdy 
\left(
\begin{array}{c}
|{\bf u}_{i,I}|^2\\
|{\bf u}_{i,C}|^2 
\end{array}
\right)
.\end{equation}
We can obtain these compressible and incompressible kinetic energy spectrum by 
considering momentum space Fourier transform, as
\begin{equation}\label{K-energy}
\left(
\begin{array}{c}
K_{i,I}\\
K_{i,C}
\end{array}
\right)= \frac{m_1}{2m_i}  
\int  dk_xdk_y \left(
\begin{array}{c}
\left|{\cal F}_{i,I}(k_x,k_y)\right|^2\\
\left|{\cal F}_{i,C}(k_x,k_y)\right|^2
\end{array}
\right)
,\end{equation}
where
\begin{align}
\left(\begin{array}{c}
\mathcal{F}_{i,I}({\bf k}) \\ \mathcal{F}_{i,C}({\bf k})
\end{array}\right)
= \frac{1}{2\pi} \int dxdy\, e^{-ik_xx-ik_yy} 
 \left(\begin{array}{c} {\bf u}_{i,I}\\  {\bf u}_{i,C}\end{array}\right).  
 \end{align}
From Eq.~(\ref{K-energy}), the total incompressible and compressible kinetic 
energies can be obtained by extending to two components the procedure 
detailed in \cite{2012Bradley}. Within this procedure, we first obtain the 
spectral density in $k-$space in polar coordinates ($k,\phi_k$), with the 
final kinetic energies by integrating on $k=\sqrt{k_x^2+k_y^2}$, as follows:
\begin{equation}
\left(
\begin{array}{c}
K_{i,I} \\
K_{i,C}
\end{array}
\right)= \int_0^\infty dk 
\left( \begin{array}{c}
{\cal K}_{i,I}(k)\\
{\cal K}_{i,C}(k)
\end{array}\right),
\end{equation}
where
\begin{equation}\label{Kinetic-k}
\left(\begin{array}{c}
{\cal K}_{i,I}(k)\\
{\cal K}_{i,C}(k)
\end{array}
\right)=
\frac{m_1 k}{2m_i}  \int_0^{2\pi}d\phi_k
\left(
\begin{array}{c}
\left|{\cal F}_{i,I}(k_x,k_y)\right|^2\\
\left|{\cal F}_{i,C}(k_x,k_y)\right|^2
\end{array}
\right)
\end{equation}
express (for both components $i$ of the mixture) the respective 
incompressible ($I$) and compressible ($C$) kinetic energy spectrum 
over the wavenumber $k$.
 
\subsection{Numerical approach}
For the numerical simulations, applied to the three different kinds of instabilities
in BEC mixtures that we are reporting, we use the split-step Crank-Nicolson method 
to solve Eq.~(\ref{2d-2c}). It is followed by performing the Fourier transforms to 
reach the compressible and incompressible kinetic energy spectra. The details for the
numerical calculations of the corresponding velocity power spectra can be obtained 
in Refs.\cite{2012Bradley,2022Bradley}.
With the ground-state solutions previously prepared in the imaginary time, the 
dynamics are followed by real-time evolutions, in which the applied external 
potential and miscibility conditions are responsible for the kinds of instabilities 
we are investigating. 
Within our dimensionless defined quantities ($l_\perp$ for space and $1/\omega_\perp$
for time), the numerical simulations are carried out by using a square grid with
400$\times$400 points, with box length $L\equiv L_{x,y}=80$
{($\Delta x=\Delta y=0.2$)}, with the corresponding wavenumber 
infrared limit being $k_L=2\pi/L$.
The time step $\Delta t$ is chosen to be $10^{-3}$. 
For each species, the respective  full-dimensional healing lengths
are assumed fixed, such that $\xi_i=\sqrt{\hbar^2/(m_i\mu_i)}\sim 0.4\; l_\perp$,  
implying ${\mu_2}=({m_1}/{m_2})\mu_1$. As we are assuming equal number of atoms 
for the two species, with identical intraspecies scattering lengths, only the 
small mass difference ($m_1$/$m_2\approx$ 0.98) plays a role in determining the 
chemical potential.
Given the above length unit, the corresponding dimensionless healing 
length is $\xi=0.4$, which is covered by approximately two grid points 
($\xi=2\Delta_x$). Consequently, the smallest physically meaningful structures, 
on the order of a few $\xi$, are resolved by multiple grid points, even at 
longer time evolution, where fine-scale structures and vortex tangles are prominent.
The corresponding \emph{infrared cutoff} wavenumber (large scale) is 
$k_L = 2\pi/80 \approx 0.0785$, while the \emph{ultraviolet} cutoff wavenumber
(small scale) imposed by the grid is $k_{\rm max}=\pi/\Delta x = \pi / 0.2 = 15.7 $.

\section{Instability simulations in binary Bose-Einstein condensates}\label{sec3}
In our present study, we investigate dynamical processes that occur with a slight 
mass-imbalanced binary mixture, represented by the $^{85}$Rb-$^{87}$Rb BEC system,
prepared initially in an immiscible space configuration. We explore phenomena 
analogous to the Kelvin-Helmholtz and Rayleigh-Taylor instabilities, 
engineered by externally applied perturbations, as well as by quenching the 
nonlinearity of the mixture from immiscible to miscible, called IMQT.
These standard instabilities typically evolve from stationary states, where infinitesimal
perturbations grow over time. This is well-studied in classical fluid dynamics and can be 
extended to quantum gases using the Bogoliubov-de Gennes (BdG) analysis, a standard 
tool for examining the onset of instabilities in quantum systems from stationary states. 
We must stress that, in our investigations of what we call instabilities along 
the text, the systems are dynamically changed, engineered to evolve out of 
equilibrium from the beginning, due to applied external forces, or nonlinear
quenching. 
Specifically, the mathematical structure of our problem 
precludes the use of BdG analysis, prompting us to investigate these instabilities 
in a time-dependent framework. The motivation for this approach relies on 
mathematical clarity and the 
observation that the resulting patterns are similar to those seen in standard cases. 
Apart from several previous studies on this matter, further motivations are 
brought from actual ongoing experiments considering RT and KH instabilities in BEC 
and quantum fluids, such as the ones reported in Refs.~\cite{2022Mukherjee,2024Rajkov}
with the observation of KH instabilities considering single-component atomic species. 
More recently, the authors of Ref.~\cite{2024Huh} are claiming the first realization of 
KH instability in inviscid fluid, with the observation of RT instability in binary quantum 
fluid being reported in Ref.~\cite{2024Geng}.

Next, we present numerical simulations leading to three kinds of instabilities in 
binary mass-imbalanced mixtures of two BEC systems, prepared in immiscible conditions.
We present our simulations for the density dynamics, compressible and incompressible
kinetic energies, and corresponding spectra. The first two cases refer to
RT and KH kinds of instabilities, obtained by external forces keeping the same immiscible 
initial conditions. In the case of  IMQT, which refers to sudden changes in the 
non-linearity of the coupled system, two kinds of initial conditions 
are investigated for the dynamics. 
Even considering that all simulations can be easily adapted to other binary atomic
systems, particularly to spinor levels of the same atom, here
we are assuming the mass-imbalanced $^{85}{\rm Rb}-^{87}$Rb binary BEC mixture.
To help us understand the instability dynamics in the time
evolution, for each one of the density panels that we are going to show, 
the corresponding phase profile is included as a twin panel at the 
right-hand-side (rhs) of the density. 
Subtle phase variations in hue indicate smooth phase gradients (no vortices). 
In the complex dynamics, among the increasing number of vortices that start to 
be generated, a specific vortex in one of the species can be verified 
as a singularity (zero density), which corresponds to an endpoint of a 
line (a line segment, having two endpoints, refers to vortex-antivortex 
occurrence) in the phase profiles. 
Correspondingly, in the other component, no vortex can be found 
at the same location: in the density plots, such a point emerges as a maximum 
(bright spot), indicating that the hole (singularity) 
generated in the density distribution of a component is filled by the other species.

For all the instabilities we have investigated, the incompressible results are 
{ closely related to the production and motion of quantized vortices, with the 
compressible parts being due to sound-wave radiation,
density fluctuations, and other dissipative effects. Both can follow the same 
scaling when strongly coupled, with energy transfer between the modes. 
However, they do not necessarily follow the same energy cascade process, as 
the kinetic compressible mode can be radiated away as sound waves or through 
other mechanisms.} { This dynamics is discussed in different studies
related to turbulence in BEC systems, such as Refs.~\cite{1997Nore,2005Kobayashi,2014White}.
For general cases of compressible turbulence, analytic support can be 
found in Ref.~\cite{2011Aluie}}
The $k^{-3}$ scaling behavior in the kinetic energy spectrum is known to arise 
from the vortex core structure. 
Conversely, the infrared regime (when $k\xi\ll 1$) arises purely 
from the configuration of the vortices and turbulence.  
The energy injected by the vortices and their interactions can be observed in  
the infrared regime, as already noticed in Ref.~\cite{2012Bradley}. 

\subsection{Rayleigh-Taylor instability in coupled BECs}
\begin{figure*}[!htbp] 
\vspace{-1cm}
\begin{center}
\includegraphics[height=23cm,width=15cm]{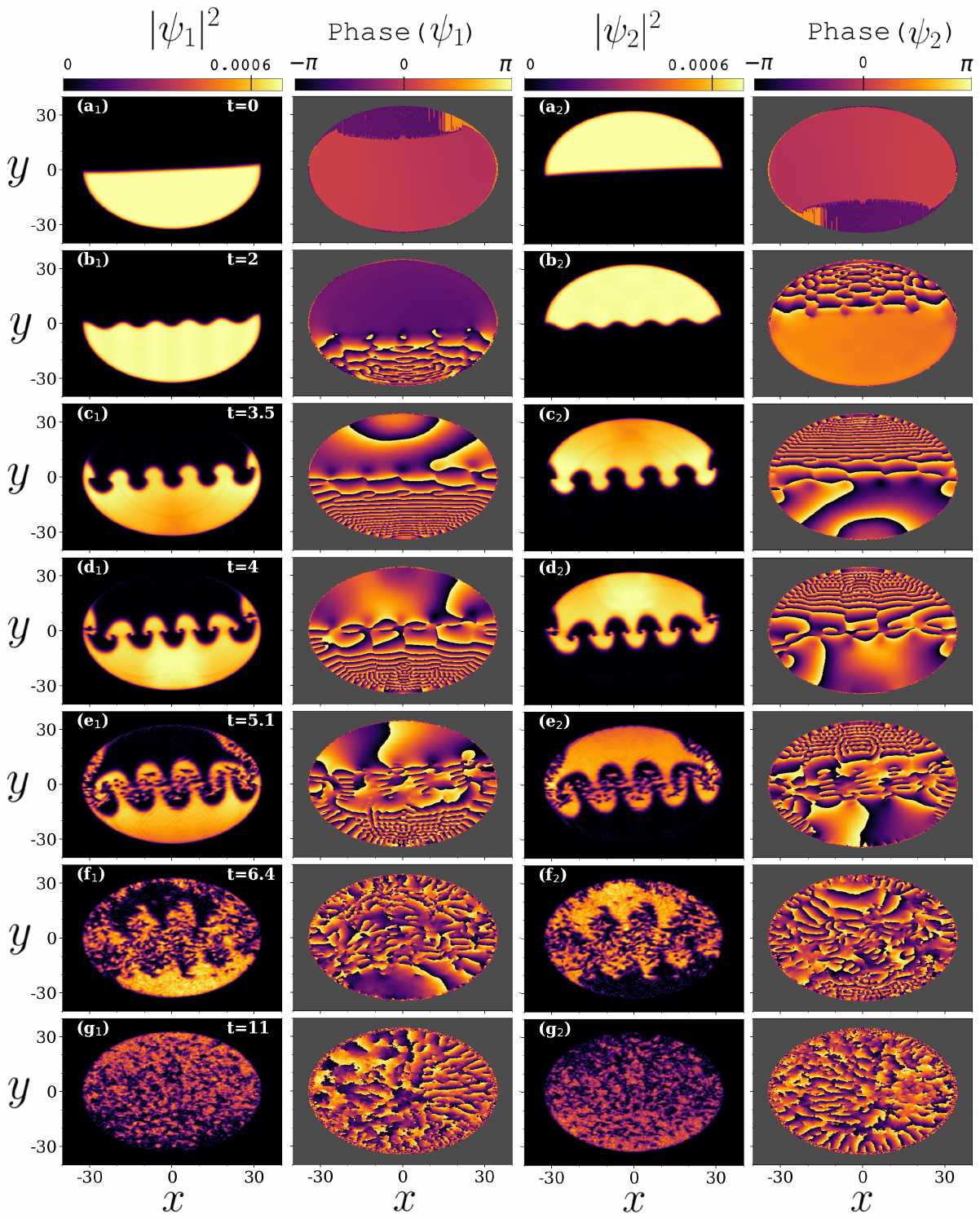}
\vspace{-1cm} \caption{(Color online)
RT instability in the binary mixture $^{85}$Rb [(a$_1$)-(g$_1$)] and $^{87}$Rb 
[(a$_2$)-(g$_2$)], shown by sample time snapshots of the densities $|\psi_i|^2$,
together with respective phases (as indicated, with $t$ given inside the panels 
for the densities). The immiscible condition $\delta=a_{12}/a_{ii}=1.05$ is 
kept along the numerical simulations. The color-bar levels for densities and 
phases are indicated at the top, with the units for time and length being, 
respectively, $\omega_\perp^{-1}$ and $l_\perp$. 
The corresponding full-dynamical evolution is provided in the 
supplemental material~\cite{Suppl}.
}
\label{fig01}
\end{center}
\end{figure*}
\begin{figure}[!htbp]
\begin{center}
\includegraphics[width=0.45\textwidth]{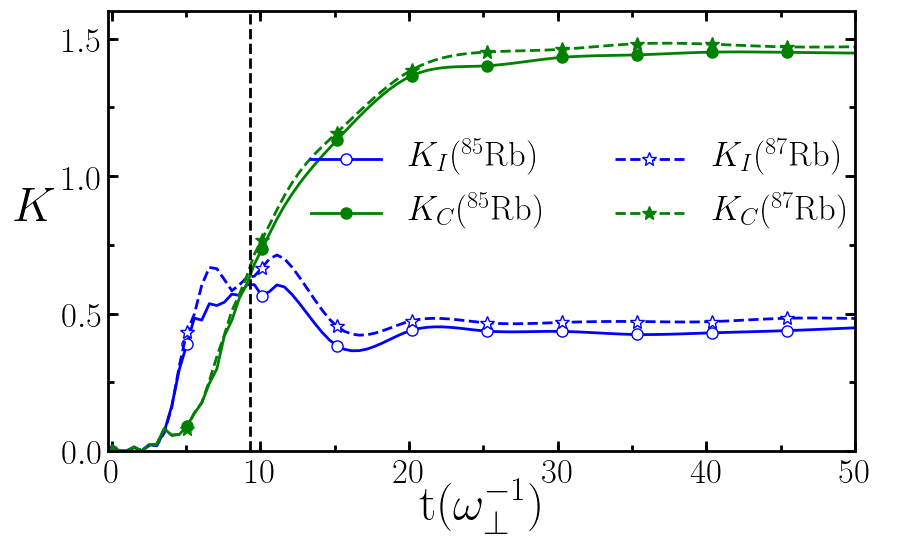} 
\caption{(Color online) Time evolutions of incompressible and compressible 
kinetic energies, $K_I$ and $K_C$ (units $\hbar \omega_{\perp}$), for the  
$^{85}$Rb-$^{87}$Rb mixture, associated with RT instabilities, with convention 
as indicated inside the panel.
The vertical dashed line at $t\approx 9$ identifies the approximate time 
instant for the energy transition from incompressible (related to the vorticity) 
to a compressible (sound waves) dominated fluid. }
\label{fig02}
\end{center} 
\end{figure}

In normal classical fluids, the RT instability occurs at the interface between 
two different-density fluids. In particular, it happens when the lighter fluid 
pushes the heavier one with the support of the gravitational force. 
To simulate the occurrence of a similar effect in ultra-cold systems, instead 
of a gravity force acting among the two species, we assume an immiscible 
binary mixture, in which the two 
elements are under the effect of external opposite linear forces acting on the two components.
The mass difference between the components can be neglected, as the simulation can 
be done as well with binary coupled systems having two different levels of the
same atom prepared in immiscible conditions.
To contemplate this model, we prepare the ground state by considering an axially separated
mixture, with the interspecies interaction chosen to be larger than the intraspecies one, 
as shown in Fig.~\ref{fig01} with $\delta=1.05$ ($a_{12}=105a_0$ and $a_{11}=a_{22}=100a_0$). 
In our simulation, the initial two-component ground-state solutions of the mixture are obtained 
by using imaginary-time calculations, immediately followed by their real-time evolution. 
In Fig.~\ref{fig01}, the upper panels (a$_1$) and (a$_2$) depict the prepared initial 
two-component
densities for the immiscible mixture at $t=0$, when starting the real-time propagation.
Due to numerical conditions when concluding the imaginary-time calculations and starting the 
real-time evolution, it happens that at $t=0$ the borderline separation 
(between the two immiscible fluids) starts slightly 
inclined as related to the horizontal line.
As anticipated, the phase profiles are included at the rhs of the density plots.
Related to the ground-state density panels (a$_1$) and (a$_2$) 
at $t=0$, the phases are zero in the regions where the homogeneous 
fluid densities are located, as verified. However, in the zero-density 
regions, the respective phases (both species) are undefined, with the shown 
results just reflecting numerical artifacts, which appear at the 
threshold when starting the real-time propagation. 
In all the other phase profile panels, the changes in the color-darkness 
correspond to phase variations of the fluid circulating around the singularities, 
which goes from $-\pi$ (black) to $\pi$ 
(bright-yellow). Interesting vortex dynamics revealed by the phase profiles
can be verified in detail by enlarging the phase-profile panels, together 
with the corresponding density panels.  
Throughout the temporal evolution, the two-body scattering lengths are kept constant.
To start the dynamical instability, the trap interaction \eqref{2Dtrap} is modified 
by a sinusoidal $x-$direction perturbation applied to the first component for a short time 
interval from $t=0$ till $t=2$,
creating the density oscillation presented in the panels (b$_1$) and (b$_2$). 
The simulation follows (for $t>2$) with the sinusoidal perturbation removed
{ from the potential, and replaced by the linearly varying perturbations 
$\nu_1 y=1.2\;y$ and $\nu_2 y=-1.2\;y$, which provides constant forces} $\nu_i$ in the 
$y-$direction. More explicitly, by using the usual Kronecker $\delta_{i,j}$
(=1, for $i=j$; 0, otherwise) and the step-function $\Theta(x)$ (=1, for $x>0$; 0, for $x<0$), 
the modified dimensionless perturbations $\widetilde{V}^{RT}_i$ can be expressed by
\begin{eqnarray}
\widetilde{V}^{RT}_i(x,y;t)&=&V_i(x,y)+\nu_i y\Theta(t-2) 
\nonumber\\&-&\delta_{i,1}\Theta(2-t)\cos\left(\frac{x}{2}\right).\label{VRT}
\end{eqnarray} 
To observe dynamics analogous to the ones of the RT instability, 
the constant forces acting on both components are in opposite $y-$directions 
to mimic a system under gravity.
However, with the external forces applied as prescribed, there is no need for
the species to have different masses; the same approach could also be applied 
to spinor states of the same atom.
The main relevant requirement for starting the dynamics is the 
immiscibility provided by the fixed relation between the inter and intraspecies
scattering lengths, which we are assuming such that $\delta=1.05$.

\begin{figure*}[!ht]
\begin{center}
\includegraphics[width=0.92\textwidth]{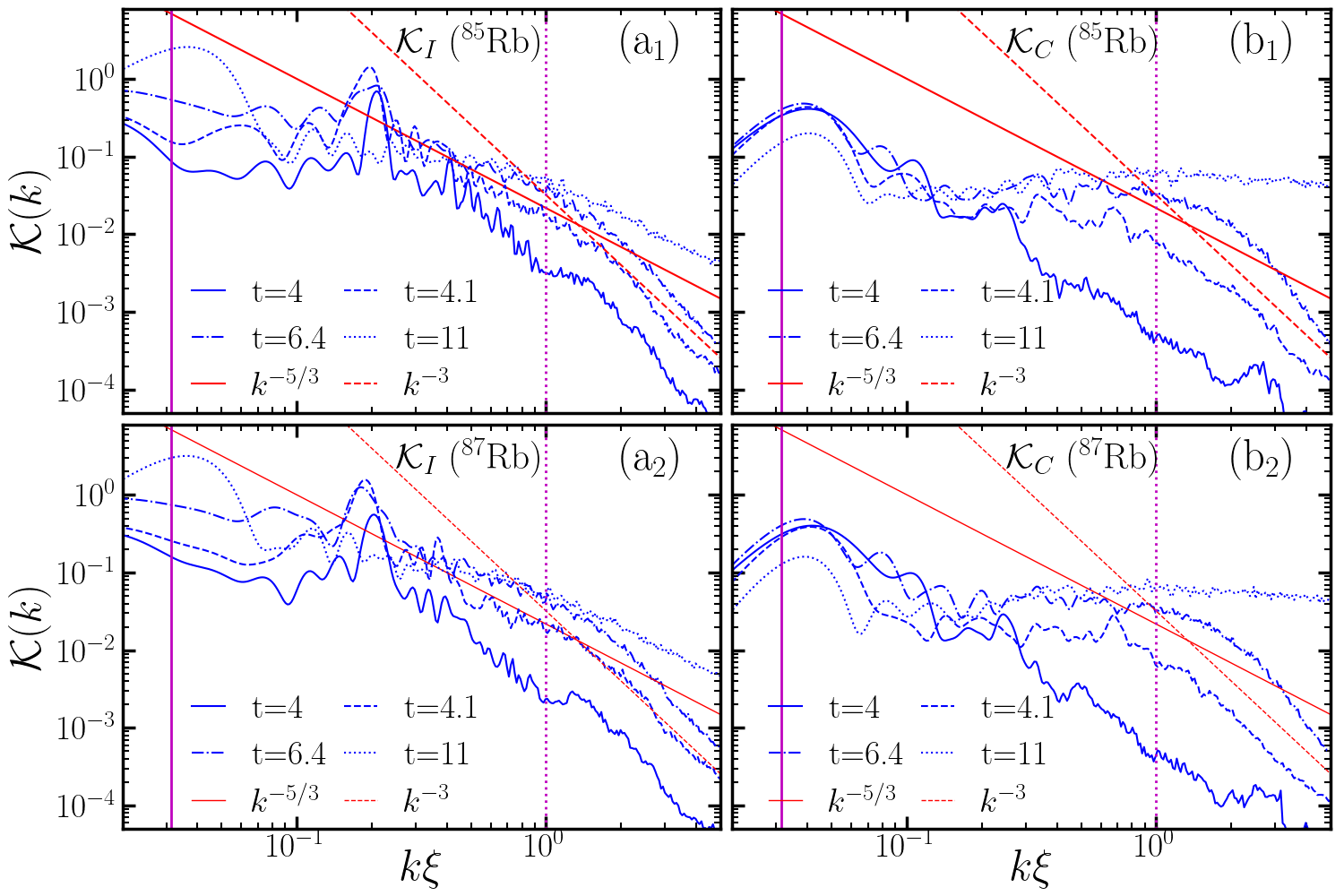}
\caption{(Color online) 
Incompressible (a$_i$) and compressible (b$_i$) kinetic energy spectra,
${\cal K}(k)$ (units of $\hbar\omega_\perp l_\perp$), for the RT dynamics, shown as 
functions of the dimensionless $k\xi$ for the $^{85}$Rb (upper panels) 
and $^{87}$Rb (lower panels) components of the mixture, considering four time 
instants $t$ in the evolution (as indicated).
The inclined straight lines provide the $k^{-5/3}$ and $k^{-3}$ behaviors, 
for comparison. The two close instants 4.0 and 4.1 refer to the fast behavior 
transition following the panels (d$_i$) shown in Fig.~\ref{fig01}.
The vertical lines refer
to the size of the box (infrared limit), $k\xi=k_L\xi=0.01\pi$ (solid line), 
and the starting ultraviolet region $k\xi=1$ (dotted line), where $\xi=0.4$.
The allowed maximum (defined by $\Delta x=0.2$)
goes to $k_{\rm max}=5\pi\sim 15.7$.
} 
\label{fig03}
\end{center} 
\end{figure*}

For the mass-imbalanced mixture, we follow a previous study done in  Ref.~\cite{Kishor2020}, 
in which the linear force is introduced by a small perturbation in the trap. 
Related experimental and theoretical proposals for applying linear 
perturbations have  also been discussed in Refs.~\cite{Sasaki2009,Bezett2010,Takeuchi2010, 
 Kobyakov2011, Kobyakov2014, 2011McCarron}. 
Quite similar to our present numerical simulation, recently it was 
reported in Ref.~\cite{2024Geng} an experimental observation of RT instability in a binary
quantum fluid composed of two hyperfine levels of $^{23}$Na, which are trapped by a 
square potential with minimal symmetry breaking at $y=0$. 

In Fig.~\ref{fig01}, the first panels (a$_i$) and (b$_i$) of the evolution 
display the density separations between the condensates, showing the  
sinusoidal perturbation, introduced in the interval $0<t\le 2$ (as given by
Eq.~\eqref{VRT}). They are helpful 
to observe the onset of RT instability in the $^{85}$Rb-$^{87}$Rb mixture.
The dynamical process in the evolution follows just after the replacement of the sinusoidal perturbation
by the external forces, at $t=2$, with mushroom pattern formations and plenty 
of vortex dipoles generated under the head of the mushrooms. 
Together with the vortex dynamics, 
large phonon production can also be noticed in the evolution of the coupled densities.
The panels (c$_i$) through (g$_i$) provide 
indicative snapshot results of the RT instability simulation. 
In panels (g$_i$), a tendency can be observed for the components of the 
immiscible mixture to occupy distinct spaces inside the trap, in opposite positions as 
compared to the original ones. Such dynamics can better be visualized in the animation 
corresponding to Fig.~\ref{fig01} provided in the supplementary material~\cite{Suppl}.

By considering the approach presented in Sect. IIB, the instability is also being 
analyzed through the kinetic energy spectrum, in Fig.~\ref{fig02}. 
The vorticity (measured by the increasing number of vortices) is primarily 
associated with the increase in the incompressible part of the kinetic energy, 
namely $K_{i,I}$, whereas the sound-wave production is related to the 
compressible part, $K_{i,C}$. The respective behaviors can be observed from
the results shown in Fig.~\ref{fig02}, for the evolution of both 
kinetic energy parts. Following the legends and caption of Fig.~\ref{fig02},
the solid-blue (dashed-blue) line refers to the incompressible kinetic energy 
$K_{^{85}{\rm Rb},I}$ ($K_{^{87}{\rm Rb},I}$), whereas the 
solid-green (dashed-green) line refers to the corresponding compressible part, 
$K_{^{85}{\rm Rb},C}$ ($K_{^{87}{\rm Rb},C}$).
As noticed, the vorticity of both components fast increases in the
interval $2 < t < 11$, when the incompressible part of the kinetic energy is 
dominating the dynamics. The sound-wave production starts to increase for 
larger times, dominating the dynamics in the long-time evolution. 
It corresponds to phonon excitations in the fluid, which
can be attributed to the energy transfer obtained from the 
production and annihilation dynamics in a continuous process that 
persists for longer times~\cite{2005Kozik,2016Mendonca}.
As the external forces are maintained along the dynamics, the $K_{i,I}$
behaviors (related to vorticities), which have maxima near $t=10$, 
remain very high for $t > 11$, besides being lower 
than the $K_{i,C}$ (related to the large sound-wave production).
Therefore, the incompressible energies $K_{i,I}$ do not decay down to zero 
for longer times, as the coupled system has been constantly driven by the 
applied forces. As the associated vortex numbers remain quite large along 
all the dynamics, in this case, the time evolution of the incompressible 
kinetic energy can better represent the vorticity.
A direct numerical association between the vortex numbers with the 
incompressible energies $K_{i,I}$ in the long-time evolution can be noticed 
for moderate instabilities, not generated by external forces, 
such as in the cases we are considering in the final part of the present study.
There, external forces are absent, implying a significant vortex number
reduction.

As indicated in Fig.~\ref{fig02}, the vortex production starts near $t\approx 3.5$ 
in the dynamics, with the corresponding energy $K_{i,I}$ (vortex production) 
being greater than $K_{i,C}$ (related to sound waves) for both components of 
the mixture until $t\approx 9$. 
Plenty of vortex dipoles are being generated during this time interval, 
with the phonon (sound-wave) contribution being less significant. 
 The complete dynamics behavior reflects the interplay between 
the nonlinear repulsive interactions given by the immiscible condition $a_{12}>a_{ii}$
with the applied external potential in both condensates, which is exerting 
a pressure between the two clouds that are colliding against each other.
 In the initial dynamics, the system is still dominated by nonlinear repulsive 
 interactions (with the incompressible kinetic energy higher than the compressible one).
However, the constant external forces act against the nonlinear repulsion, so that
the coupled system undergoes a transition at some point (indicated by our results 
to being close to $t\sim 9$). 
Even considering that the coupled gas maintains the nonlinear conditions of immiscibility
along the simulation, the constant external forces end up effectively dominating the 
dynamics, causing pressure perturbations in the densities that propagate due to 
compressibility. The results given in Fig.~\ref{fig02} provide an analysis through the
kinetic energy spectrum of the interacting density dynamics shown in 
Fig.~\ref{fig01}, reflecting dynamics sharing some analogy with colliding classical 
fluids.
The vertical dashed line in Fig.~\ref{fig02} indicates approximately the time position 
at which the transition occurs, from the dominance of incompressible kinetic energies
(for $t\le 9$) to compressible ones (for $t\ge 9$). 
Consistently, it is noticeable that the more massive element has the components of 
the kinetic energy slightly greater than those obtained for the less massive one.
The differences are enhanced particularly in the case of incompressible energies 
close to the time when the transition happens from incompressible to compressible dominance. 

The kinetic energy spectra over the wavenumber $k$ can provide a better approach 
for analyzing the instabilities and turbulent behaviors that can occur in the evolution. 
For that, we have Fig.~\ref{fig03}, which shows results related to the dynamics presented 
in Figs.~\ref{fig01} and \ref{fig02}, considering four instants of interest in the onset 
of the instability. So, in Fig.~\ref{fig03}, we are displaying in log scales the 
corresponding spectral functions, ${\cal K}_{i,I}(k)$ and ${\cal K}_{i,C}(k)$, given 
by  Eq.~(\ref{Kinetic-k}), 
as functions of $k\xi$, where $\xi$ refers to the assumed common healing lengths of both 
elements. The inclined straight lines inside the panels are just guidelines showing the 
expected classical scaling behaviors for turbulence, $k^{-5/3}$ (red-solid line) and 
$k^{-3}$ (red-dotted line), corresponding to the behavior that goes to the  
infrared ($k\xi\ll 1$) and ultraviolet ($k\xi\gg 1$) regions, respectively. 
The solid vertical line in the infrared regime, at $k=k_L=2\pi/L$,  
provides the size of the box, with the dotted vertical line indicating the position 
where $k=1/\xi$.
These scalings are useful in identifying possible time intervals at which the 
dynamics may follow more closely the classical scalings for turbulence, 
such that some similarities can be traced between classical and quantum behaviors.

As shown in the four panels of Fig.~\ref{fig03}, at which we consider four instant
sample results in the onset of instability, 
the classically predicted scaling behavior $k^{-3}$ for the ultraviolet regime
(for $k\xi\gg 1$) can be recognized approximately
close to the time interval between 4 and 6.4 (units $1/\omega_\perp$), for both
incompressible and compressible kinetic energies.
This interval is consistent with the results shown in Fig.~\ref{fig01} for the 
dynamical evolution of the densities. Deviations are noticed 
outside this interval, particularly for larger times of ${\cal K}_{i,C}(k)$ results.
By going to the ultraviolet limit, the spectrum starts to become
more flattened, as noticed in the compressible results. This
behavior, which can be verified for $t>9$, is represented in Fig.~\ref{fig03} 
by the dotted line for $t=11$. Also noticed in Fig.~\ref{fig02}, for 
$t>9$, we start having dominance of compressible effects 
(sound waves and density fluctuations) in the dynamics.
On the intermediate $k$ region, for $0.2<k\xi< 1$, we can approximately identify 
the $k^{-5/3}$ behavior only in case of the incompressible kinetic energy results,
(see, the results  for $t=4.1$ and 6.4), with the results for $t=4.0$ and 4.1 
indicating a transition behavior from $k^{-2}$ to $k^{-5/3}$.
Essentially, the incompressible results are dominated by the motion of 
quantized vortices, in the initial time interval. However, in this case,
the compressible energies are not being transferred through a cascade process as the
incompressible ones, which is understood due to sound-wave radiation and 
dissipative effects.
\begin{figure*}[!htbp]
\vspace{-1cm}
\begin{center}
\includegraphics[height=23cm,width=15cm]{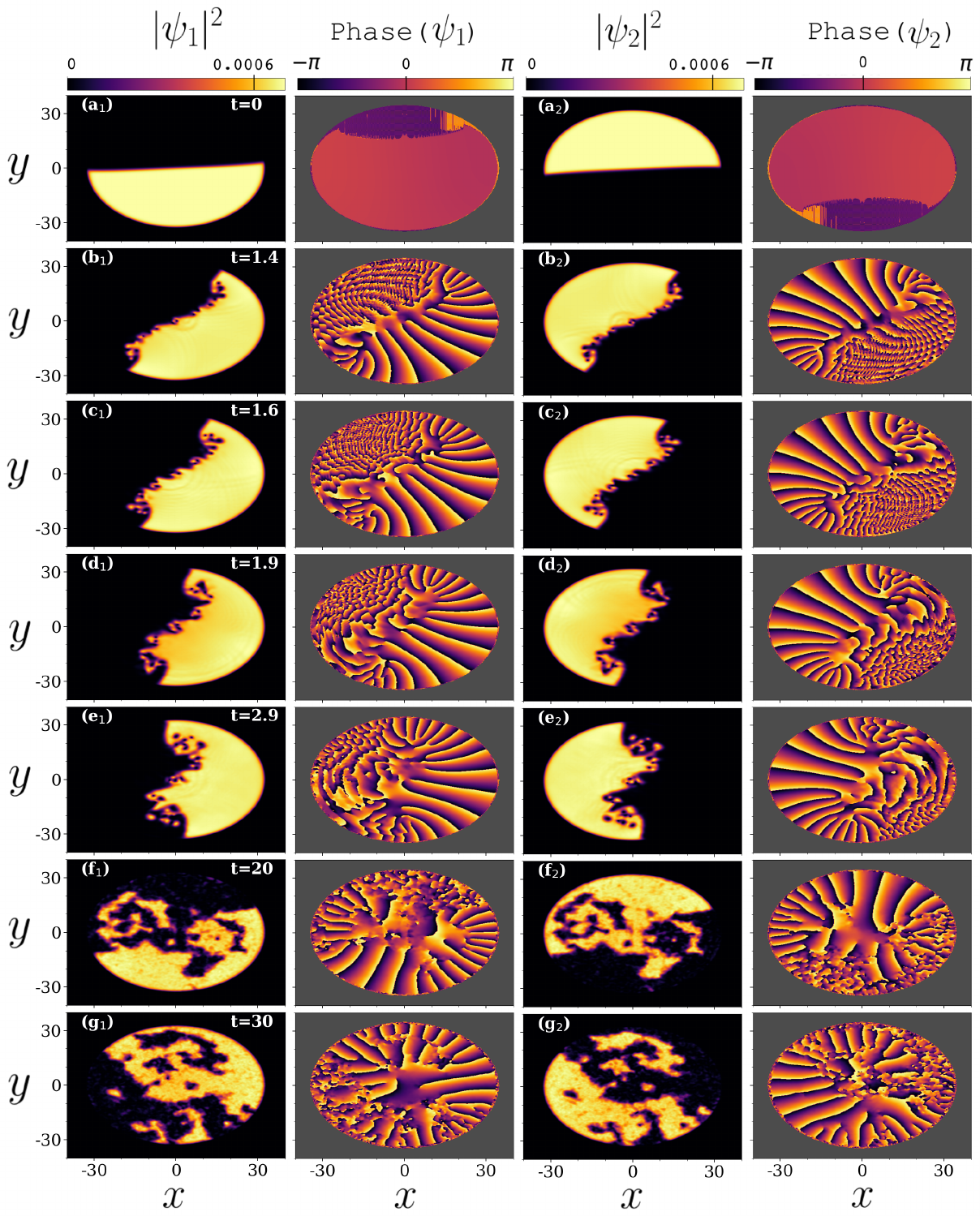}
\vspace{-1cm}\caption{ (Color online)
KH instability in the binary mixture $^{85}$Rb [panels (a$_1$)-(g$_1$)] and 
$^{87}$Rb [panels (a$_2$)-(g$_2$)], shown by sample time snapshots (with $t$ 
given inside the left panels) of the respective densities $|\psi_i|^2$ and 
phases, obtained by numerical simulations with the immiscible condition 
$\delta=1.05$. Here, a constant linear force $\nu_i=(-)^{i+1}0.7$ (in the 
$x-$direction) is applied to the components. With the color-bar levels for 
densities and phases indicated at the top, the units for time and length 
are, respectively,  $\omega_\perp^{-1}$ and $l_\perp$.
The corresponding full-dynamical evolution is provided in the 
supplemental material~\cite{Suppl}.}
\label{fig04}
\end{center}
\end{figure*}

\begin{figure}[!htbp]
\begin{center}
\includegraphics[width=0.48\textwidth]{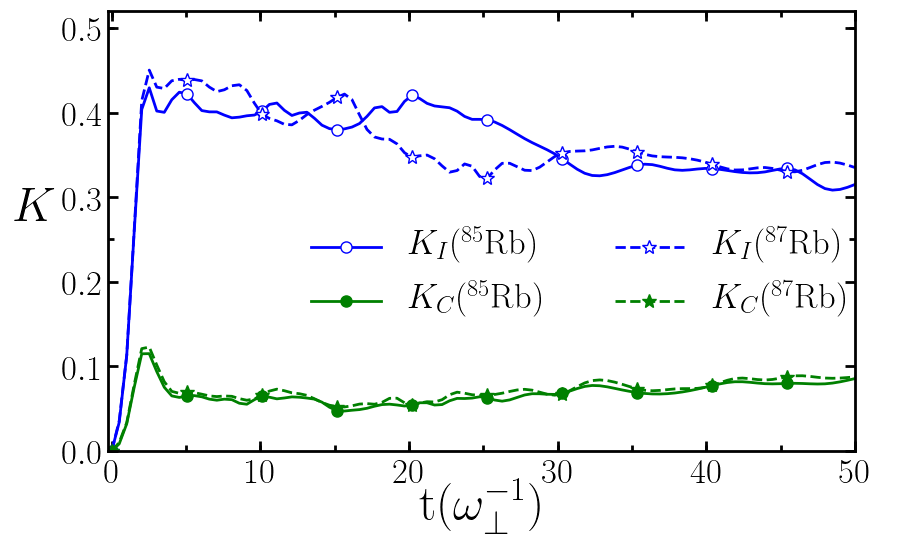}
\caption{(Color online) Time evolution of the incompressible (empty symbols) and
compressible (filled symbols) kinetic energies $K$ (units $\hbar \omega_{\perp}$) 
of the two components, evidencing KH instabilities.  
The incompressible results become larger than the compressible ones due 
to the dominance of vortex emission with their interaction. As legends indicate, the 
solid lines are for $^{85}$Rb, with dashed ones for $^{87}$Rb results.  }
\label{fig05}
\end{center} 
\end{figure}

\begin{figure*}[!htbp]
\begin{center}
\includegraphics[width=0.95\textwidth]{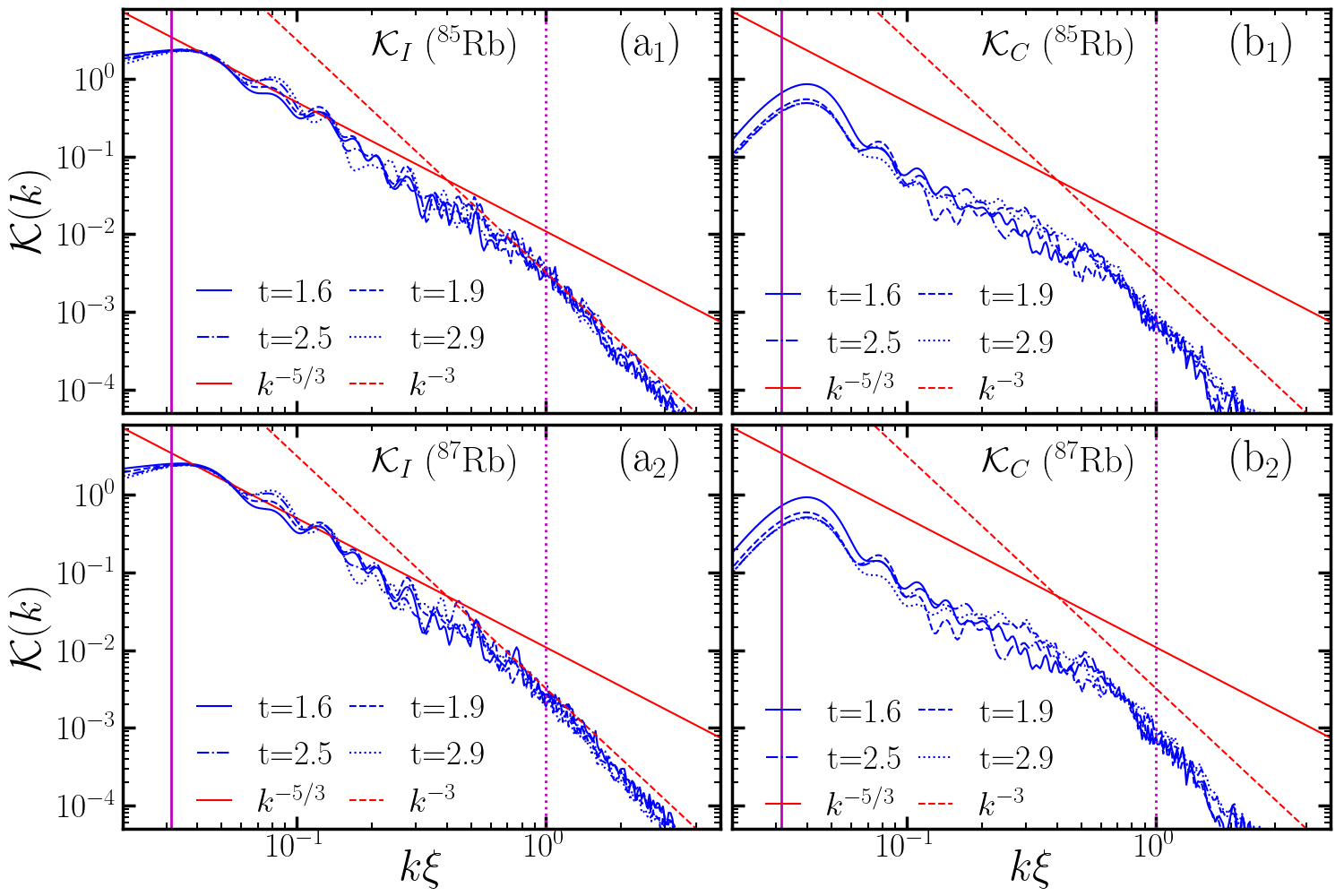}
\caption{(Color online)
Related to the KH instability represented in Figs.~\ref{fig04} and \ref{fig05}, 
it is shown the incompressible [panels (a$_i$)] and compressible [panels (b$_i$)] 
kinetic energy spectra, ${\cal K}(k)$ (units of $\hbar\omega_\perp l_\perp$), 
as functions of $k\xi$ for the  first ($^{85}$Rb) (upper panels) and second ($^{87}$Rb) 
(lower panels) components of the mixture,
considering four different time instants $t$ in the evolution (as indicated inside the panels).
The line conventions, units, and definitions follow the same as given in the caption 
of Fig~\ref{fig03}.}
\label{fig06}
\end{center} 
\end{figure*}

\subsection{Kelvin-Helmholtz instability in coupled BECs}
The KH instability occurs due to velocity differences across the interface in the classical 
two-fluid system. The KH instability significantly influences the topology of the interface 
between different-density fluids. This can be developed by introducing a velocity difference 
between the immiscible mixtures. 
On quantum KH instability, some recent experiments and analyses were reported in 
Ref.~\cite{2024Huh}.
In our present numerical simulations for the onset of KH instability, we are assuming
the phase-separated $^{85}$Rb\,-$^{87}$Rb BEC mixture. The ground state is prepared as 
in the case we have used for RT instability [See, panels (a$_i$) in Fig.~\ref{fig04}], 
such that it will help us to analyze the main differences in both dynamics. 
To obtain an effective velocity difference between the species across the interface, in 
this case, the linear interactions are applied to both components, in opposite $x-$directions, 
along all the real-time dynamics. The applied constant forces are such that 
$\nu_1=0.7$ for the first component ($^{85}$Rb), and $\nu_2=-0.7$ for the second 
component ($^{87}$Rb), with the modified dimensionless potential 
$\widetilde{V}^{KH}_i$ given by
{\small\begin{equation}
\widetilde{V}^{KH}_i(x,y;t)=V_i(x,y)+\nu_i x.
\end{equation} 
}The instability can be observed in Fig.~\ref{fig04} by a
series of panels for the coupled mixture along the time evolution.
As in the RT case, for all the density panels, we 
are showing the corresponding phases as twin panels in the 
right-hand side of the densities. With the two species being prepared
in the same way as in the RT case, at $t=0$ the phases are zero 
in the regions where the homogeneous densities are located, being 
undefined in the zero-density region. Here, for the KH instabilities,
by an examination of the phase profiles along the dynamics, we can observe
that the occurrence of vorticity is stronger near the interface 
between high- and low-density regions, as well as inside the
low-density region, which can be seen more clearly by enlarging 
the phase-profile panels.
One should notice, at $t=1.4$, the fluid motion going to the right in 
(b$_1$), and to the left in (b$_2$). Due to the radial confinement and 
the initial configuration of the two condensates, the two clouds move 
in opposite directions, with velocity initially given by $\pm v_x$ 
($v_y=0)$. Once reaching the curved surface of the confinement, the 
elements of the fluid going right receive a non-zero velocity upward, 
$+v_y$; with the other fluid receiving a velocity component downward, $-v_y$. 
Together with the immiscibility of the two species, the net effect 
on the two clouds is given by a rotation of the full coupled
system. As the configuration geometry changes over time, 
with the forces kept in the same opposite directions for both 
condensed clouds, the full system starts to be fractalized, as 
shown in the panel of Fig.~\ref{fig04}. 
Roll-up structures forming vortices can be observed, 
which are generated by producing vortices with the same signs. This vortex production 
can be detected by using a package for vortex distribution studies, available 
in~\cite{2018Bradley}. 
The vortex production and distribution are maintained in the evolution for times 
much longer than in the case of RT instability, as noticed by comparing Figs.~\ref{fig04}
and~\ref{fig01}. In this respect, the different dynamics observed for the RT and KH 
instabilities 
can be examined from the respective evolutions of the incompressible and 
compressible kinetic energies. In the case of KH instability, corresponding to
Fig.~\ref{fig04}, Fig.~\ref{fig05} shows that the vorticity dominates all the
evolution dynamics, with the incompressible kinetic energies for both components, being 
more than three times larger than the corresponding compressible kinetic energies
(which are related to sound-wave propagation).
The KH dynamics is quite different from that observed for RT instability.
A common feature they share is in the respective results for the incompressible 
energy, which do not decrease to zero, as both systems are driven by external forces.
In the case of RT instability, the vorticity dominance occurs only for $t<11$.
This behavior relies on the fact that the constant perturbation is initially introduced 
along the surface separating the immiscible fluid, which causes both components 
to move to opposite borders of the trap. As the forces continue acting in such 
immiscible coupled fluid, one of the immiscible components tries to occupy spaces 
not occupied by the other component in an almost permanent movement.

Also, in this case, as related to the KH instability dynamics shown in 
Figs.~\ref{fig04} and \ref{fig05}, to help us look for similarities with 
corresponding classical behaviors for turbulence, we have the
Fig.~\ref{fig06} with the corresponding results for the incompressible (left panels) 
and compressible (right panels) kinetic energy spectra over the wavenumber $k$. 
As shown, both elements, in this case, have quite similar spectral distributions
(incompressible and compressible), as functions of $k\xi<1$, practically not 
varying as time evolves along the period of instabilities,
represented four time instants, $t=$1.6, 1.9, 2.5, and 2.9$\omega_\perp^{-1}$, as indicated. 
In the ultraviolet region, the results follow more consistently the $k^{-3}$ 
behavior, being noticed that such behavior starts already 
near $k\xi\sim 0.5$. However, going to smaller values of $k$,
the Kolmogorov behavior, $k^{-5/3}$, is being followed more closely in the
interval $k_L<k<0.5/\xi$. Both compressible and incompressible energies
have similar trend behavior, with the incompressible kinetic energy behavior 
deviating slightly down, such that the cascade change from $k^{-5/3}$ to 
$k^{-3}$ is less pronounced.
It is known that the compressible energy spectrum may follow the same 
Kolmogorov-like scaling $k^{-5/3}$ as for the incompressible one, if the
two modes are strongly coupled~\cite{2023Barenghi,2012Bradley}.
This is observed in the KH instability results given in Fig.~\ref{fig06},
clarifying that the density oscillations and other compressible effects
responsible for the compressible effects are not so strong to modify
the behavior. 
Also, in this case, the similar scalings for both incompressible and compressible 
modes indicate that the external forces are exciting equally such modes.
Comparatively, we can see a different behavior
in the compressible kinetic energy of RT, shown in Fig.~\ref{fig03}, in which stronger 
oscillations are observed at all different instability instants that we have
selected. The expected turbulent-like behavior is quite restricted in time, for the
RT instabilities, mainly due to the strong dominance of compressible effects, 
as shown in Fig.~\ref{fig02} and panels (b$_i$) of Fig.~\ref{fig03}.

\subsubsection*{\bf The RT and KH instabilities and the miscibility}
\begin{figure}[!htbp]
\begin{center}
\includegraphics[width=0.47\textwidth]{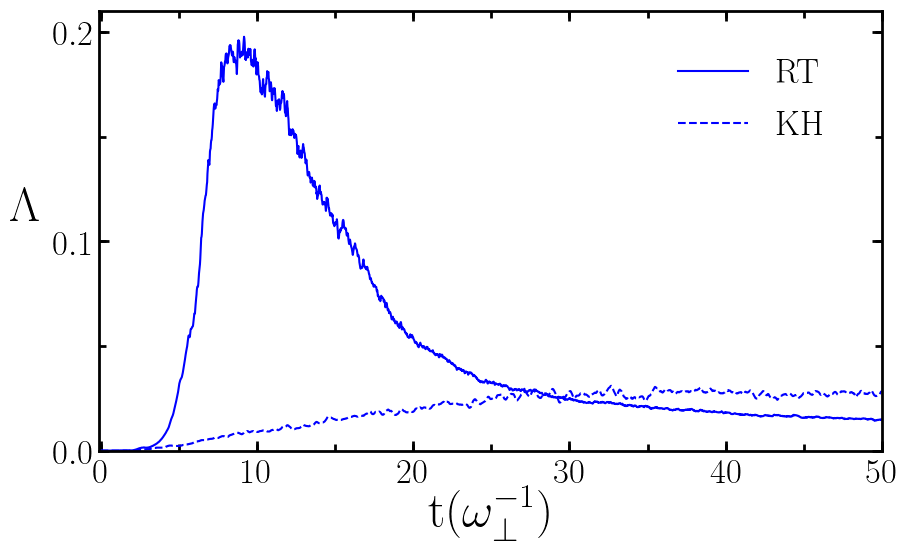}
\caption{Time evolution of $\Lambda$ (dimensionless) [Eq.~(\ref{eta_new})], representing
the density overlaps for the RT (solid line) and KH (dashed line) instabilities  
in the $^{85}$Rb\,-$^{87}$Rb mixture. Respectively, the density evolutions are  
in Figs.~\ref{fig01} and \ref{fig04}). 
}
\label{fig07}
\end{center}
\end{figure}

Concluding the analysis of the RT and KH instabilities, discussed in subsections 
A and B above, with Fig.~\ref{fig07} we are verifying how the initial immiscibility 
regime of the mixture behaves throughout the temporal evolution, considering 
the density overlap parameter $\Lambda$ defined by Eq.~(\ref{eta_new}).
As observed, because the inter-species and intra-species interactions are kept the same 
along the process of evolution, the mixture remains almost immiscible, 
besides the attractive force applied between the two species, 
which makes both components of the mixture share overlapping regions.
In the case when we have RT instability, there is a tendency for the system to become more 
miscible during the time interval when the main instability is being observed, 
with a maximum for $\Lambda$ near $t=9$ (close to the time instant when compressible 
effects, leading to stronger density oscillations, start dominating the dynamics, 
as seen in Fig.~\ref{fig02}). 
Still, the overlap represented by $\Lambda$ remains below 20\%, next decreasing below 
2\% in the long-time evolution. For the case of dynamics reminiscent of the
KH instability, with the density evolution of the mixture represented by 
Fig.~\ref{fig04}, $\Lambda$ increases slightly from zero and becomes stable near 3\% for 
longer-time simulations.
Therefore, the Fig.~\ref{fig07} results show that, for both RT and KH cases 
produced by the linear-force perturbation, there are no immiscible to 
miscible transitions. The tendency of the RT instability to become more miscible occurs only
in a shorter time interval when one can notice a transition in the kinetic energy spectrum,
from incompressible to compressible kinetic energy dominance.
\begin{figure*}[!htbp]
\vspace{-2cm}
\begin{center}
\includegraphics[height=23cm,width=15cm]{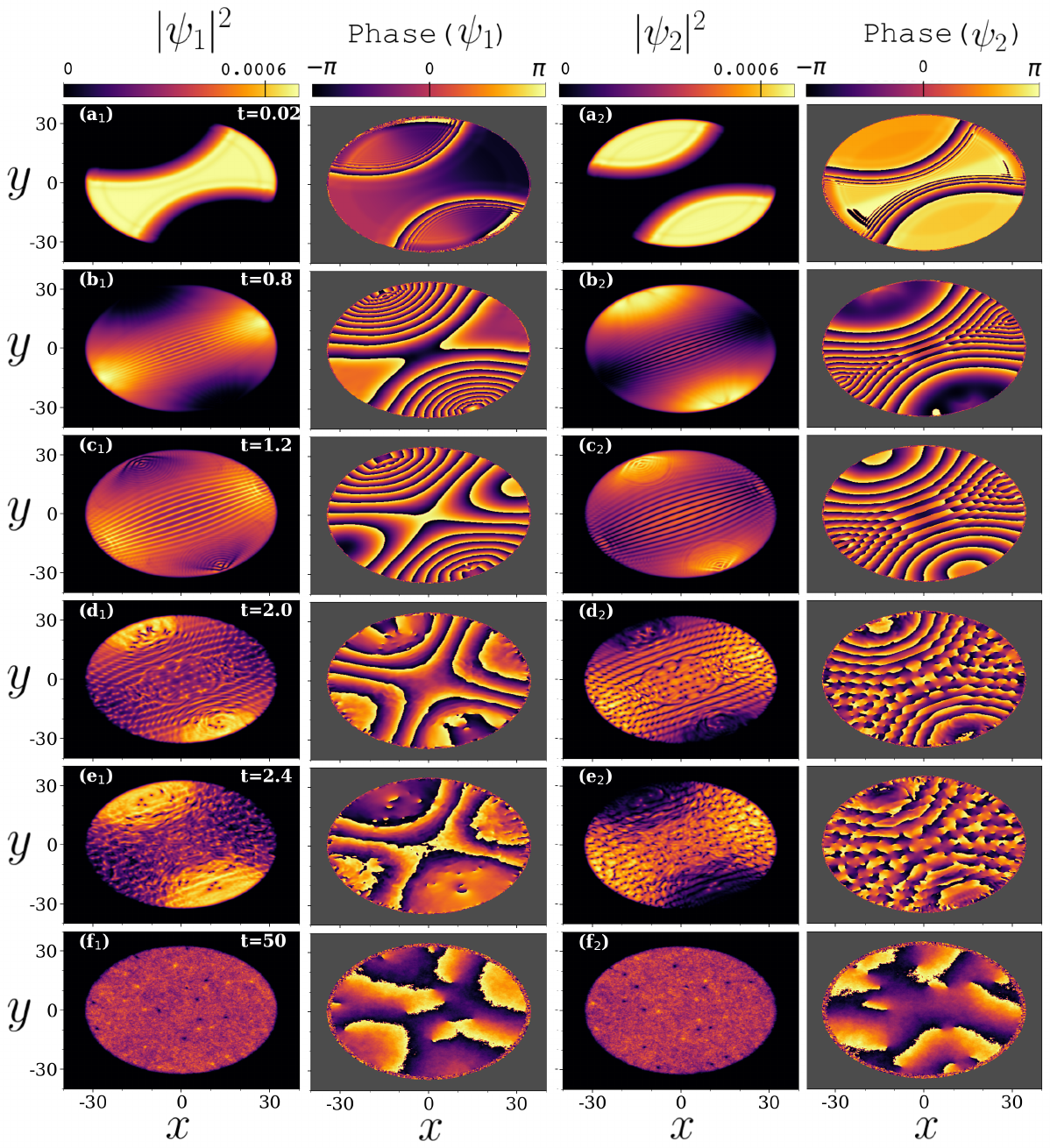}
\vspace{-2cm}\caption{(Color online) 
 IMQT instability for the $^{85}$Rb\,-$^{87}$Rb mixture is shown through the evolution of 
 the densities and phases. The ground state is prepared in an immiscible regime (with 
 $\delta=1.02$) in a projected ``tennis-ball" shape configuration, with the $^{85}$Rb 
 centrally located [(a$_1$)]  and $^{87}$Rb in the remaining trap confinement [(a$_2$)]. 
 The evolution starts with a sudden reduction of $a_{12}$, going to $\delta=0.75$, 
which remains along the dynamics. The snapshot instants $t$ are indicated inside the 
density panels. The units for time and length are, respectively, $\omega_\perp^{-1}$ 
and $l_\perp$. The corresponding full-dynamical evolution is provided in the 
supplemental material~\cite{Suppl}.
}
\label{fig08}
\end{center}
\end{figure*} 

\begin{figure}[!htbp]
\begin{center}
\includegraphics[width=0.48\textwidth]{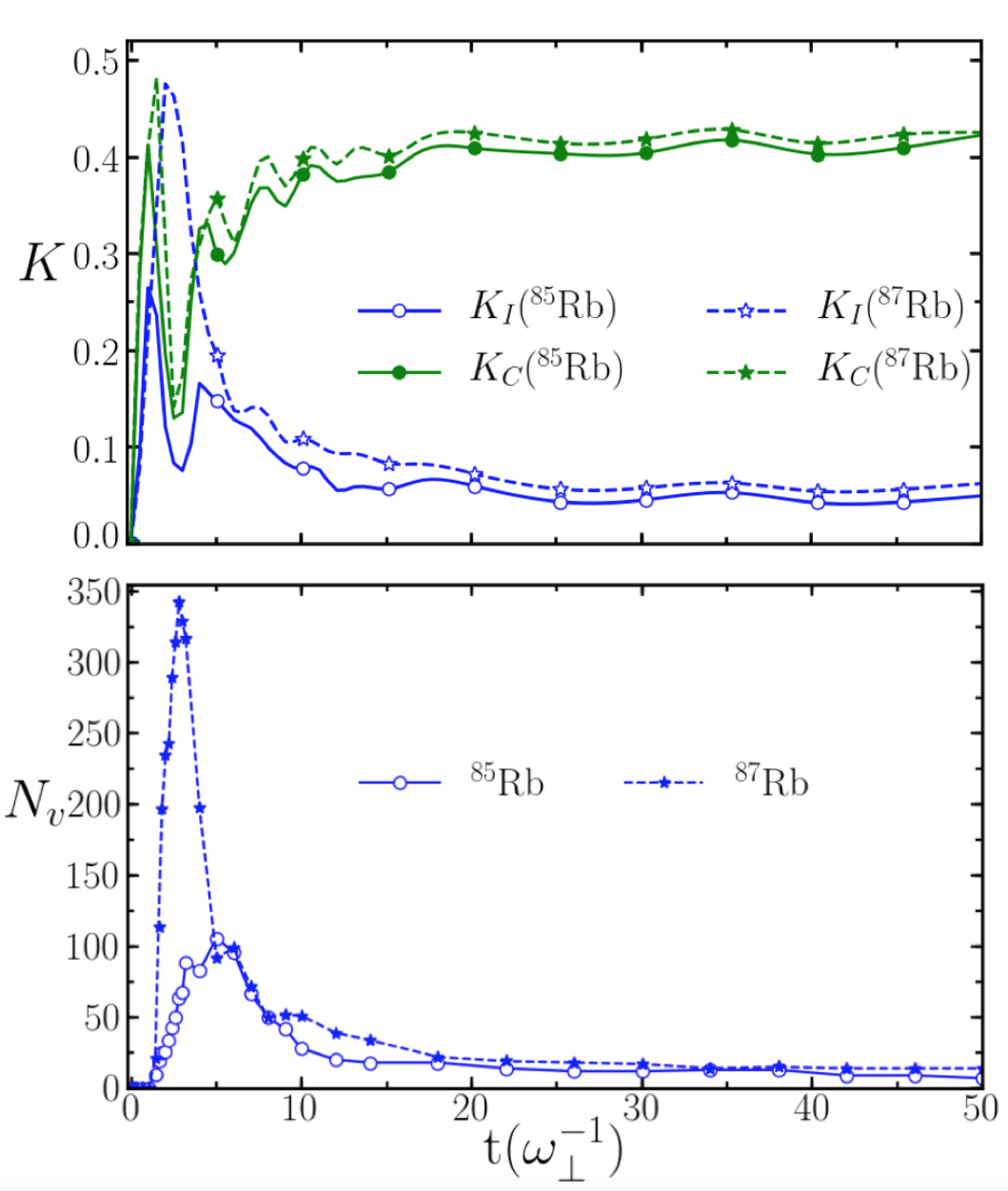}
\caption{(Color online)
For the IMQT instability given in Fig.~\ref{fig08}, with the first species, $^{85}$Rb,
initially, at the central part, it is shown the time evolutions of the compressible and 
incompressible kinetic energies $K$ (units $\hbar \omega_{\perp}$) (upper frame) and 
the corresponding number of vortices $N_v$ (lower frame), with convention as indicated 
inside the frames.}
\label{fig09}
\end{center} 
\end{figure}

\begin{figure*}[!htbp]
\begin{center}
\includegraphics[width=0.95\textwidth]{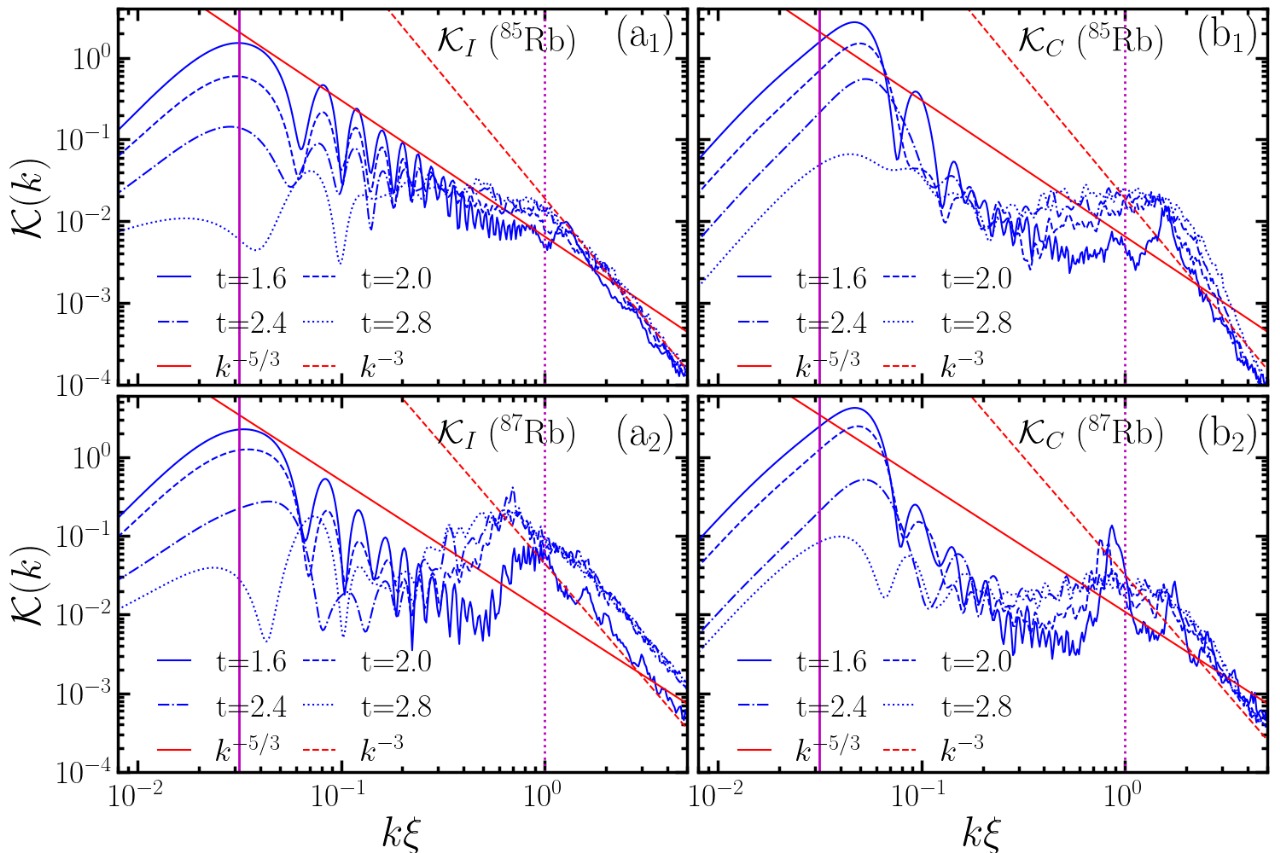}
\caption{(Color online)
Related to IMQT instability represented in Figs.~\ref{fig08} and \ref{fig09}, 
it is shown the incompressible [panels (a$_i$)] and compressible [panels (b$_i$)] 
kinetic energy spectra, ${\cal K}(k)$ (units of $\hbar\omega_\perp l_\perp$), 
as functions of $k\xi$ for the  first ($^{85}$Rb) (upper panels) and second ($^{87}$Rb) 
(lower panels) components of the mixture,
considering four different time instants $t$ in the evolution (as indicated inside the panels).
The line conventions, units, and definitions follow the same as given in the caption 
of Fig~\ref{fig03}.}
\label{fig10}
\end{center} 
\end{figure*}

\subsection{Immiscible to miscible quenching transition instability in coupled BECs}
The following simulations with the binary $^{85}$Rb\,-$^{87}$Rb mixture consider the 
dynamical IMQT instability by preparing the original coupled system in an immiscible 
condition ($\delta>1$), with the interspecies two-body interaction larger than the 
intraspecies one, $a_{12}>a_{ii}$ (also here,  $a_{11}=a_{22}$). 
Different from our simulations for RT and KH instabilities, here we are not 
applying linear perturbations to obtain the dynamical evolution.
Instead, a quench-induced transition is applied by introducing a sudden reduction
in the two-body interspecies scattering length, such that the coupled system 
moves from immiscible ($\delta>1$) to miscible ($\delta<1$) conditions.
The motivation for studying the dynamic behavior of cases where instability 
arises from nonlinear interactions is to compare it with the previous studies 
in which the onset of instabilities was carried out through external forces. 
Therefore, we apply the same approach as before. Here, we find it appropriate to 
probe two different initial immiscible 2D spatial configurations for the ground 
state, as follows: 
(i) First, with {\it tennis-ball projected format}, 
with one species at the central part, identified as ``central"; 
(ii) Second, with the species side-by-side, identified as ``axial". 

\subsubsection*{\bf IMQT with ``tennis-ball" shaped initial state}
The first initial configuration considered in our study of IMQT instability 
is by having the two species within a tennis-ball 2D projected shape, within a 
three-sliced initial configuration, having the $^{85}$Rb in the central part, 
with the other species, $^{87}$Rb, located at both sides of the centrally 
localized component. 
This configuration is shown just after the start of the dynamics by the 
panels (a$_1$) and (a$_2$) of Fig.~\ref{fig08}, for the densities 
together with respective phases.
The quench-induced interaction is introduced by a sudden reduction of 
$a_{12}$, such that the miscibility of the mixture goes from $\delta=1.02$ 
(with $a_{12}=102a_0$, $a_{ii}=100a_0$)
to $\delta=0.75$ (with $a_{12}=75a_0$).
Here we are assuming a value of $\delta=1.02$ slightly smaller than 
$\delta=1.05$, which we are going to assume in the other IMQT simulation, 
considering that we would like to explore further the effect of a slightly 
different $\delta$ in the observed initial interference fringes of the 
densities to be discussed.

Given the initial configuration and quenching, the miscibility starts to occur 
from both sides of the centrally located component, at the two spatial borders 
shared by the two species. The dynamics can be followed through the snapshots 
(a$_i$) to (f$_i$) for the time evolution of the densities 
with corresponding phase profiles, shown in Fig.~\ref{fig08}. 
The respective kinetic energy evolutions, with the associated vortex dynamics,
and spectral analyses follow, respectively, through Figs.~\ref{fig09} and \ref{fig10}.
The onset of the instability dynamics can be observed in the overlap of both 
densities, in the initial transient period till $t\approx 2.4$ [as shown in the 
panels (b$_i$) to (e$_i$) of Fig.~\ref{fig08}], when going from the immiscible 
to a miscible configuration, with nonlinear interference patterns being noticed.
Particularly enhanced by the phase profiles, the formation and propagation of 
dark solitons in the fluid can be associated with these patterns. Along the
dynamics, we can observe the breaking of these soliton structures in 
vortex-antivortex pairs. In this regard, see panel (d$_2$) as an example, 
where vortex-antivortex pairs are generated near the center of the trap.
Correspondingly, at the same locations, bright dots can be observed
in the other component of the fluid, as shown in the density panel (d$_1$).
This indicates that species 1 fills the hole left by species 2.

Similar to the ones experimentally observed in Ref.~\cite{1997Andrews} for the 
interference of two Bose condensates, these patterns are attributed  
to the difference between the initial and final miscibility factor $\delta$, 
which in this case corresponds to a quenching reduction of the
repulsive interspecies parameter $a_{12}$ of about 27$a_0$. 
The connection between the observed interferences obtained in the GP 
mean-field theory with dark solitons (which can be generated for repulsive 
nonlinear interactions) was discussed in Ref.~\cite{2010Frantzeskakis}. 
The interference patterns noticed in Fig.~\ref{fig08} resemble planar dark solitons 
in propagation through the binary mixture. 
Among several related studies of dark solitons in BEC we can mention the  
Refs.~\cite{1995Pelinovsky,2006El,2012Yan}. 
Planar dark solitons are subject to snake instability, where
the dark solitons decay into vortex dipoles~\cite{1996Tikhonenko}.
These vortex dipoles are indeed observed in our simulation.
{ For fermionic superfluids, snake instability of dark solitons was
also studied in Ref.~\cite{2013Cetoli}, within the Bogoliubov-de Gennes theory of 
the Bose-Einstein condensate (BEC) to BCS crossover.}
The particular geometric form of the interference fringes shown in Fig.~\ref{fig08} 
in the evolution of the miscibility also occurs due to the initial condition of
the prepared immiscible coupled states, with $^{87}$Rb fragmented in two 
distinct regions. With the sudden reduction in the interspecies repulsive
interaction, the $^{87}$Rb atom species move towards the center from both sides,
interacting with the $^{85}$Rb located in the central region, as seen 
in the panels (c$_i$) of Fig.~\ref{fig08}.
Plenty of vortex dipoles and sound waves (phonon excitations)
are produced in both densities, as one can observe in the corresponding time evolution,
which are supported by the results shown in Fig.~\ref{fig09} 
for the incompressible (related to vorticity) and compressible (related to sound 
waves) parts of the kinetic energy spectrum. 
The vortices are spontaneously generated when the components are interacting,
leading to annihilations and sound-wave production. 
As noticed from the two panels of Fig.~\ref{fig09} there is a close relation 
between the incompressible part 
of the kinetic energy with the vortex numbers $N_v$, in the evolution of the 
mixture, with the peaks of $N_v$ slightly shifted to the right of the 
corresponding peaks observed for the $K_{i,I}$ ($i=^{85}$Rb,$^{87}$Rb). 

In the onset of instabilities, near $t\sim 5$, one can also
observe a strong peak that occurs in the results for the $^{87}$Rb density
(element 2), much larger than the maximum obtained for the other element, $^{85}$Rb.
This result can be explained, considering that the component $^{87}$Rb is
initially located outside the center, such that the intermediate space 
(occupied by the $^{85}$Rb) works as an effective barrier inside the 
$^{87}$Rb condensate (as a double well). As discussed in 
Refs.~\cite{2010Wen,2017Sabari}, 
one cannot neglect the hidden vortices inside this internal low-density region,
because they carry angular momentum and are essential to satisfy the 
Feynman's rule of vortices~\cite{1955Feynman}. As the coupled system is 
under an immiscible to miscible configuration, the number of vortices 
(for both species) converge to similar results for longer times.

In the long-time evolution of the mixture, we notice that 
sound-wave propagations due to phonon excitations become dominant, as noticed 
in the upper panel of Fig.~\ref{fig09}. $K_{i,I}$ and $N_v$, representing the
vorticity of both species, decrease to more or less permanent stable limits. 
In this asymptotic limit, one can follow the 
dynamics of the vortex propagation inside the coupled fluid. The particular
vortices observed in the densities of one of the components can be followed 
in the numerical simulations, which are represented by holes in movement 
inside the density profile. 
Correspondingly, one can observe a density increase of the other species at the 
same positions, implying one component fills the spatial holes opened by the 
other component inside the trap. Related to this, we notice
that a similar effect has been reported in Ref.~\cite{2024Caldara} when characterizing 
superfluid KH instability of a fluid in the presence of a second component. 
Such long-time dynamics can still be distinguished in both coupled fluids, 
since they are still in a condition not fully miscible, with $\delta=0.75$. 
This will be further discussed, considering the 
asymptotic incomplete overlap $\Lambda$ of the densities.

The spectral behavior of the incompressible and compressible kinetic 
energies, respectively given by ${\cal K}_I(k)$ and ${\cal K}_C(k)$,  
can be analyzed through the results shown in Fig.~\ref{fig10} for the
two components, in which the turbulent dynamics are being identified 
in the initial period of the evolution by considering four instants. 
As noticed, the classical Kolmogorov $k^{-5/3}$ 
can be evidenced approximately for $k\xi<0.5$ (by averaging the 
oscillations) only at some particular time interval close to 
$t\approx 1.6$, being clearer in case of incompressible kinetic 
energies. In the ultraviolet region, the $k^{-3}$ behavior can also
be approximately distinguished at a short time interval $t<2.8$, 
when the onset of instabilities occurs. For 
longer times, the classical spectral behavior does not occur anymore. 
The remaining vorticity and density fluctuations in the evolution can 
be followed by the incompressible and compressible energy results
shown in Fig.~\ref{fig09}, as well as, visually, by the evolution of 
the coupled densities shown in \ref{fig08}.
For all selected time instants of the dynamics, shown in this 
Fig.~\ref{fig10}, transient energy increases (incompressible and compressible) 
are also noticed occurring in an intermediate $k$ interval, 
starting near $k\xi\approx 0.8$, when the classical behavior, expected 
at least for the incompressible part, should change from $k^{-5/3}$ to $k^{-3}$. 
This kind of effect, which is deviating from the expected Kolmogorov's 
cascade transition in the energy spectra, can be associated with the strong 
oscillations between the coupled species.
Due to nonlinear dynamical interactions at intermediate scales, 
with energies temporarily accumulated before being fully transferred 
between the scales, the effect can also be recognized as similar to the quantum 
superfluid 3D bottleneck effect, 
at which energy piles up at scales just before the dissipation 
range~\cite{2007Lvov,2004Connaughton}. 
Often seen in 3D turbulence, such an effect can also occur in 2D or quasi-2D 
systems. Within a 2D approach, the authors of Ref.~\cite{2009Bos} have 
pointed out that such an effect is not only observable at small scales, 
but can also be artificially created at large scales. 
Within the perspective of further studies, one can also notice that 
such an effect is more pronounced in the results obtained for the 
element not initially at the center, which can be associated with the 
existence of two surface borders separating the species.

 \begin{figure*}[!htbp]
\vspace{-2cm}
\begin{center}
\includegraphics[height=23cm,width=15cm]{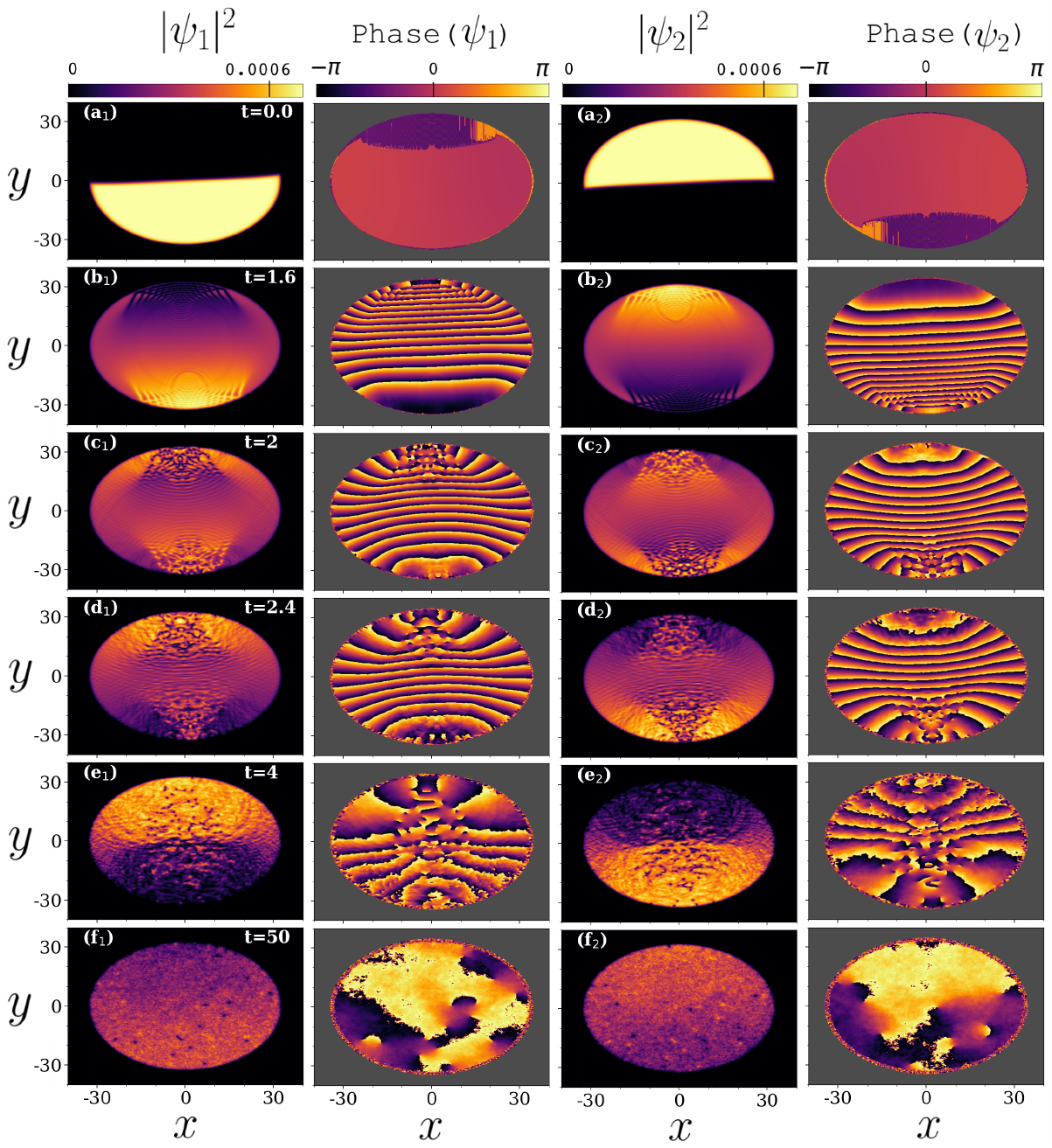}
\vspace{-2cm}\caption{(Color online) 
 IMQT instability for the $^{85}$Rb\,-$^{87}$Rb mixture
is shown through the evolution of the densities and phases. The ground state is prepared in an 
 immiscible regime (with $\delta=1.05$) in an axial geometrical configuration, 
 with $^{85}$Rb located in the lower part [shown in (a$_1$)] 
and $^{87}$Rb in the upper part [shown in (a$_2$)].
The evolution starts with a sudden reduction of $a_{12}$, going to $\delta=0.75$, 
which remains along the dynamics.
The snapshot instants $t$ are indicated inside the left panels.
The units for time and length are, respectively,  $\omega_\perp^{-1}$ and $l_\perp$. 
(Among the supplemental material~\cite{Suppl}), we show the corresponding full-dynamical
evolution.)}
\label{fig11}
\end{center} 
\end{figure*} 
   
\begin{figure}[!htbp]
\begin{center}
\includegraphics[width=0.48\textwidth]{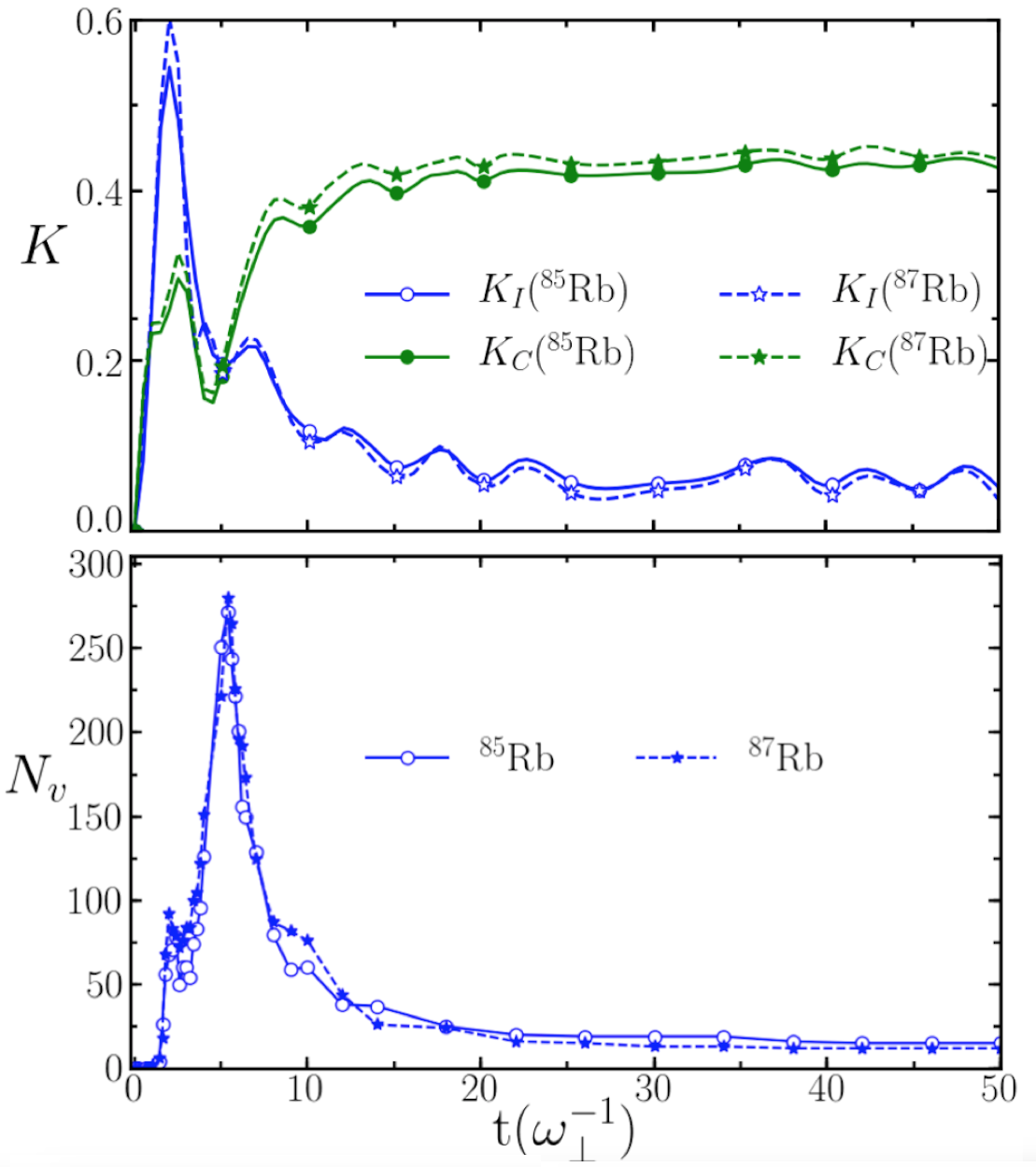}
\caption{(Color online)
For the IMQT instability given in Fig.~\ref{fig11} with axial
initially separated elements, $^{85}$Rb (solid lines) and $^{87}$Rb (dashed lines), 
the above results are for time evolutions of the kinetic energies $K$ 
(units $\hbar \omega_{\perp}$) (upper frame) and the corresponding number of 
vortices $N_v$ (lower frame), with convention as indicated inside the frames.
}
\label{fig12}
\end{center} 
\end{figure}

\begin{figure*}[!htbp]
\begin{center}
\includegraphics[width=0.95\textwidth]{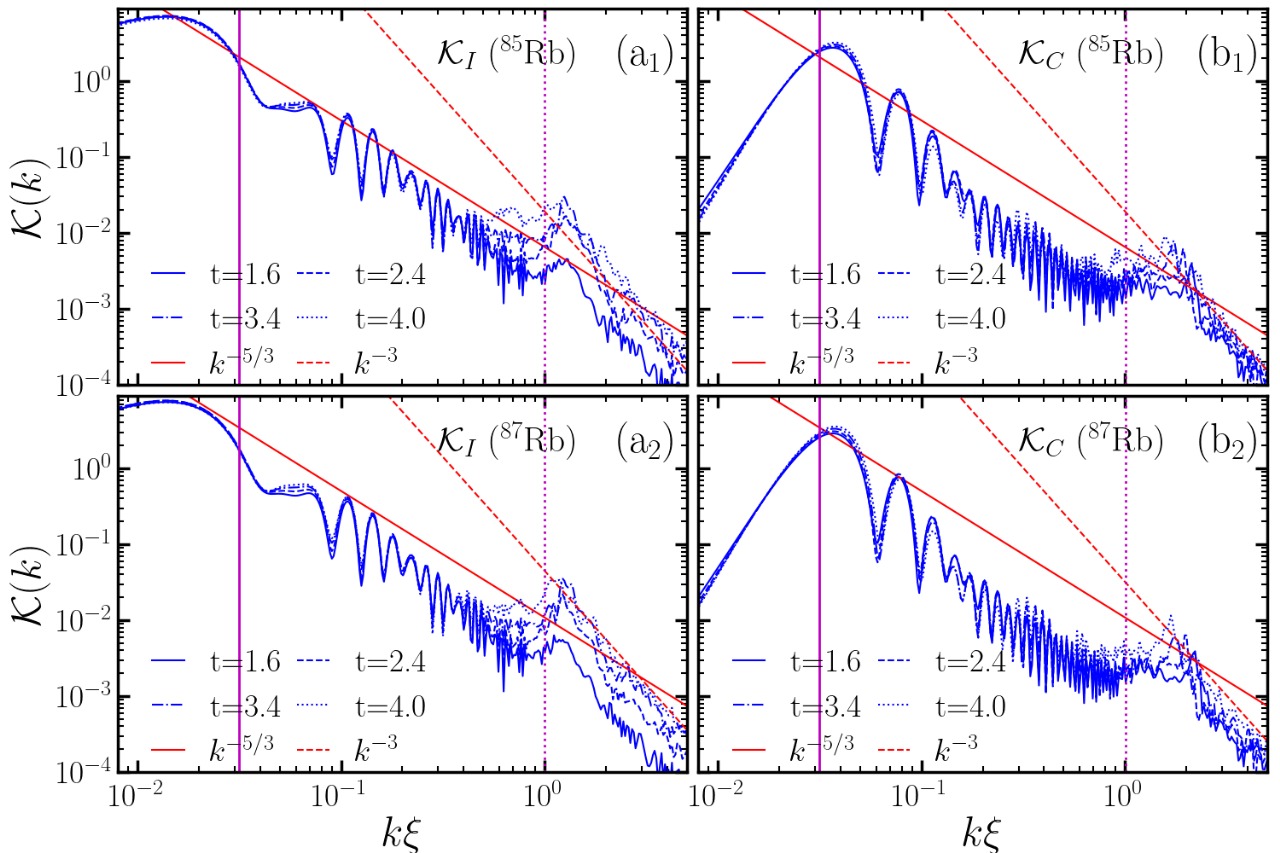}
\caption{(Color online)
Related to the IMQT results given in Figs.~\ref{fig11} and \ref{fig12}, 
the incompressible [panels (a$_i$)] and compressible [panels (b$_i$)] kinetic energy spectra,
${\cal K}(k)$ (units of $\hbar\omega_\perp l_\perp$), 
are shown as functions of $k\xi$ for the first ($^{85}$Rb) (upper panels) 
and second ($^{87}$Rb) (lower panels) components of the mixture, considering four different
time instants $t$ in the evolution (as indicated inside the panels).
The line conventions, units, and definitions follow the same as given in the caption 
of Fig~\ref{fig03}.
}
\label{fig13}
\end{center} 
\end{figure*}

\subsubsection*{\bf IMQT with axial shaped initial state}
\noindent The other initial configuration considered for the IMQT instability 
is represented by both immiscible densities located side-by-side, as shown 
in panels (a$_1$) and (a$_2$) of Fig.~\ref{fig11},
with the time evolution of densities and corresponding phase profiles
represented by the other panels of this figure.
In this case, the initial immiscible condition ($\delta=1.05$) and spatial 
configuration (for the ground state prepared in imaginary time) are
the same as in the cases of Figs.~\ref{fig01} and \ref{fig04}.
For each density panel along the evolution, we are showing the corresponding 
phase profile as a twin panel on the rhs of the respective density. 
Related to the density panels (a$_1$) and 
(a$_2$) at $t=0$, the ground state shows zero phase, with the observed
break in the uniformity related to random noise in the
threshold when starting the real-time propagation.
As previously explained, when discussing the initial condition
of the simulations shown by Fig~\ref{fig08}, here the ground state is 
initially prepared with $a_{12}=105a_0$ ($\delta=1.05$).
With a different value for the starting immiscibility, we aim to 
explore the effect of a slightly different $\delta$ in the observed initial 
interference fringes of the densities, as the systems evolve to miscible
configurations. In this regard, see panels (c$_i$) and (d$_i$) of both 
Figs.~\ref{fig08} and ~\ref{fig11}]. 
By comparing the results obtained with the two different
initial spatial configurations, in this Fig.~\ref{fig11}, we notice a more 
symmetric formation of interference patterns as the mixture evolves, 
than the ones observed in Fig.~\ref{fig08}. The vortex-antivortex
pair formations start to occur near the extreme borders of the trap, 
instead of at the center.

Figure~\ref{fig12} shows the time evolution of the kinetic energy (compressible
and incompressible) and vortex number detected, for both species. These results
should be compared with the ones given in Fig.~\ref{fig09}
(when the species 1 is centrally located), to observe the effect of the initial
conditions (spatial configuration and initial value of $a_{12}$).
In the present case, the sudden quenching of $a_{12}$ again will be to 
$a_{12}=75a_0$, such that we will have the same miscible $\delta=0.75$ 
along the dynamical evolution. As shown, the final miscible configuration 
is similar in both cases, as shown in the respective panels (f${_i}$) of 
both Figs.~\ref{fig08} and \ref{fig11}. However, the dynamical process 
noticed in Fig.~\ref{fig11} can be distinguished from the one observed in 
Fig.\ref{fig08}, in the same initial time interval of instabilities. 
This happens because, with the spatial configuration of Fig~\ref{fig11}, both 
species are symmetrically occupying the trap region. 
In the two panels of Fig.~\ref{fig12}, a similar behavior can be observed 
as in Fig.~\ref{fig09} in the long-time evolution, with the dominance of 
the compressible modes. The main difference occurs in the 
initial evolution, $t\lesssim 6$. Given the symmetric spatial distribution at $t=0$, 
the $K_{i,I}$ and $N_v$ peaks associated with the vorticities are located at the 
same time position for both components. The peaks for $N_v$ are slightly 
shifted relative to the $K_{i,I}$ peaks, as the vortex dynamics follow
the incompressible energy behavior.
In the long-term interval, with the condensate mixture searching for
miscible configuration equilibrium, the vortex numbers of both components 
reduce to about the same number below 20. 
In this time interval, both parts of the kinetic energy converge to 
different asymptotic limits, with the incompressible part being reduced 
to less than 1/8 of the compressible part. This behavior reflects the dominance 
of density fluctuations and sound-wave production, with reduced vortex dynamics, 
as one can better appreciate in the corresponding animations (See supplementary 
material).

In the four panels of Fig.~\ref{fig13} we are showing the results for the
incompressible and compressible kinetic energy spectra over the product of 
the wavenumber $k$ with the healing length $\xi$. 
Quite illustrative of the common behavior of the two components of the mixture, 
when they start with similar space configurations, 
in their initial condition [see panels (a$_i$) of Fig.~\ref{fig11}], are the 
results observed in the spontaneous 
production of vortices, shown in Fig.~\ref{fig12}. In this case, the small 
mass difference between the species only appears in the long time 
interval, with the smaller-mass component, $^{85}$Rb, corresponding 
to a slightly greater number of vortices than the ones generated in 
the $^{85}$Rb. These results follow the ones obtained by the incompressible 
and compressible modes of the kinetic energy observed in Fig.~\ref{fig12}, 
in which in the long-time interval we have more production of sound waves, 
within a behavior similar to the case of the density distribution evolution 
of Fig.~\ref{fig08}, which was discussed before.

When in the initial condition, we have the light 
element located in the sliced central part (seen in Fig.~\ref{fig08}),
the time evolution of the spectrum (shown by the $k$ behavior in Fig.~\ref{fig10}) 
presents fluctuations stronger than the corresponding case, Fig.~\ref{fig11}, 
with elements axially located at $t=0$. 
In  Fig.~\ref{fig10}, the $k^{-5/3}$ behavior can
be observed in the incompressible spectrum only at some particular time of the instability, 
with the compressible modes related to sound waves having stronger and nonuniform behavior, 
which affects the incompressible mode.
As related to Fig.~\ref{fig11}, with axial spatial configuration, the spectrum shown by
Fig.~\ref{fig13} becomes quite stable in the interval $k\xi<1$, as time evolves.  
{For both species, the approximate behavior of $k^{-5/3}$ is observed in the interval $k\xi<1$,
although the cascading behavior $k^{-3}$ has not been clearly characterized for $k\xi>1$.}
The strongest changes in the dynamical behavior of the 
spectrum, observed in Fig.~\ref{fig10}, can be attributed to 
the fact that the two densities in the miscible dynamical process have the 
atomic interspecies interactions happening in two regional borders 
near the center of the trap. In the axially separated case, shown in Fig.~\ref{fig11}, 
the dynamics is mainly dictated by the interspecies interactions starting in just one 
border separation between the species, in a more symmetric form.
   
Similarly to the case discussed for Fig~\ref{fig10}, { in an intermediate 
range of $k$, close to $k\xi\approx 0.8$ shown in Fig~\ref{fig13}, transient increases 
can also be observed in the compressible and incompressible components of the 
energy. As already discussed, this effect looks similar to the 3D bottleneck 
effect~\cite{2004Connaughton,2007Lvov}.}
Since both cases are related to sudden nonlinear changes in the interactions, such an 
effect confirms that it should be interpreted as due to nonlinear dynamical interactions 
at intermediate scales, with the energies temporarily accumulated before being fully 
transferred across the scales.
   
\begin{figure}[!htbp]
\begin{center}
\includegraphics[width=0.45\textwidth]{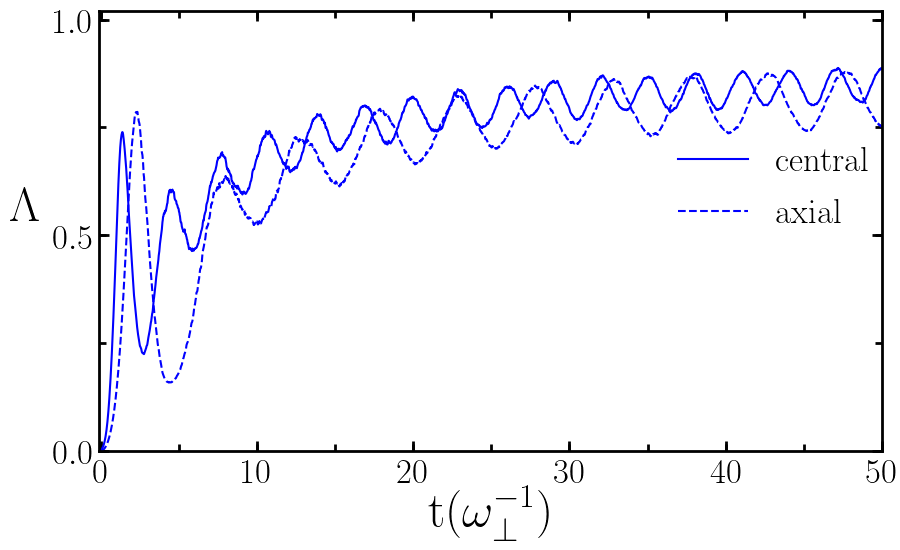}
\caption{Time evolution of $\Lambda$ (dimensionless) [Eq.~(\ref{eta_new})], representing
the density overlaps for the IMQT instabilities, with central (solid line) and 
axial (dashed line) initially space separated $^{85}$Rb and $^{87}$Rb components. 
Respectively, the density evolutions are shown in Figs.~\ref{fig08} and \ref{fig11}). }
\label{fig14}
\end{center} 
\end{figure}
 
\subsubsection*{\bf The IMQT instabilities and the miscibility}
For these two cases of IMQT instabilities, the real-time propagation of 
the overlap between the two densities is provided by the time dependence of 
$\Lambda$ [given in \eqref{eta_new}], with the results displayed in Fig.~\ref{fig14}. 
$\Lambda$ shows how dynamically the miscibility evolves, when the two-body interspecies is 
suddenly reduced, such that $\delta$ goes from 1.02 to 0.75 in case of the coupled 
densities are represented by Fig.~\ref{fig08}; and goes from 1.05 to 0.75 in case they 
are represented by Fig.~\ref{fig11}. The effect of the initial conditions in the long-time 
evolution can better be appreciated by the results shown in Fig.~\ref{fig14}, as 
indicating how the two kinds of mixtures go dynamically from immiscible to miscible 
configurations due to the sudden changes in the value of $\delta$.
In the first time interval, $t<10$, the differences are recognized as due to the distinct
spatial configurations when the sudden reduction was applied to the interspecies interaction.
In the other extreme, asymptotically, for both cases, the averaging values of the time 
evolutions of $\Lambda$ converge consistently to about the same value near 0.8, which is 
quite close to 0.75, the final quenched value of $\delta$. Such behavior, as well as the 
results obtained in both simulations shown in Figs.~\ref{fig08} and \ref{fig11}, 
indicates that, in the asymptotic limit what remains from the initial 
conditions are the averaged value of $\Lambda$ (close to 0.8 for both cases) and the 
oscillating behaviors. Apparently, from the results of Figs.~\ref{fig08} and 
\ref{fig11}, for $t>20$, the vortex dynamics are very similar in both cases, such that the 
initial spatial configurations have limited relevance for the behavior in the 
asymptotic limit. However, a striking difference between the two cases occurs not in the
averaged limit of $\lambda$, but in the different oscillating cycles of the miscibility 
overlap $\Lambda$, which we interpret as reflecting the initial sudden change of $\delta$. 
In the first case, with initial three sliced regions, named 
{\it central} (considering that the species 1 is at the central slice), the $\delta$ 
quenching varies from 1.02 to 0.75, such that $\Delta\delta_C=0.27$, whereas in the second
case, when the confining region was split into two parts, $\Delta\delta_A=0.30$. 
The central case implies a slightly faster transition, which results in a larger frequency 
(and smaller amplitude) than the second case. Another point to consider is that, 
with the initial configuration given by Fig.~\ref{fig08}, the main region for the 
interspecies interaction and possible interferences is located at the central part 
[see panels (c$_i$) and (d$_i$) of Fig.~\ref{fig08}]. 
In the case of Fig.~\ref{fig11}, the interactions occur symmetrically on both sides, 
with densities increasingly closer to the trap limits, where one can verify the main 
interference patterns [see panels (c$_i$) and (d$_i$) of Fig.~\ref{fig11}].
{By a close observation of the miscibility parameter results shown in Fig.~\ref{fig14}, 
comparing both coupled systems, brought from different initial immiscible conditions
(central, with $\delta=1.02$ and axial, with $\delta=1.05$) to the same miscible 
configurations with $\delta=0.75$,
focusing in the time interval when both coupled systems are expected to keep only the 
residual main characteristics of the quenching transition, such as in the time interval 
between $t=20$ and $t=50$,}
we have 10 oscillations in case of 
central initial configuration, against 6 oscillations in the case of axial configuration.
So, we can extract the asymptotic ratio limit between the oscillating 
frequencies of $\Lambda$, which is given by ${\cal R}_{\Lambda}\approx 5/3$ 
(5 cycles of the central case corresponding to 3 cycles of the axial case).
By looking for a relation between the asymptotic oscillating frequencies 
with the initial quenching condition, we can observe that such oscillation may 
be induced by the quenching differences, which is $\Delta\delta_A=0.30$ for the axial 
case, and $\Delta\delta_C=0.27$ for the central case. 
The ratio of these two quantities, $\Delta\delta_A/\Delta\delta_C= 10/9$, turns out to
be identical to $(2/3){\cal R}_{\Lambda}$. This may be a curious coincidence when
considering that, initially, the central case has three regions for interspecies 
interactions, whereas the axial case has only two regions. 
Indeed, it looks like an interesting topic for further investigations to consider
long-time evolutions of binary mixtures under different time-dependent transitions 
from immiscible to miscible regimes. To this aim, a time-dependent interspecies 
scattering length could be assumed as in Ref.~\cite{2024Lima} to study possible
resonant patterns in the BEC mixture.

\section{Conclusions and Summary}\label{sec4}
In this work, we have systematically investigated three distinct 
types of instabilities, each with its unique characteristics and 
underlying mechanisms. The onsets of Rayleigh-Taylor (RT) and 
Kelvin-Helmholtz (KH) instabilities are driven by linear
perturbations, consistent with their classical counterparts. In
contrast, the immiscible-to-miscible quenching transition (IMQT) 
instabilities arise from nonlinearity changes, triggered by sudden
reductions in the two-body interspecies scattering length $a_{12}$. 
These IMQT instabilities were explored under two distinct initial
conditions, highlighting the critical role of nonlinear dynamics in 
their evolution.
All numerical simulations were conducted for coupled condensates 
initially confined in a uniform 2D circular box,
prepared in immiscible configurations. Our findings not only deepen
the understanding of these instabilities in quantum systems but also
provide a foundation for further exploration of nonlinear phenomena
in multi-component Bose-Einstein condensates. Results of this study
suggest dedicated investigations for each kind of instability, 
by exploring extensions to more complex geometries, different initial
conditions, or by looking for new insights into the interplay between 
linear and nonlinear instabilities in quantum fluids.\\

\noindent 
{\bf RT dynamics:} 
Within an immiscible state, kept constant by $\delta=1.05$, with both species
axially separated, this instability is initiated 
by an oscillatory perturbation applied to one of the components for 
$t\le 2$. Subsequently, constant forces are applied to both species,
perpendicularly to the border interface between the species, in opposite 
directions throughout the process, driving the 
development of the instability. The combination of the initial
perturbation and the opposing forces leads to the characteristic 
interpenetration and mixing patterns associated with the RT 
instability, highlighting the role of competing forces in 
destabilizing the system.

With the main results provided in Figs.~\ref{fig01}-\ref{fig03},
the onset of RT instability presents a striking similarity with standard results 
obtained with a mixture of two classical immiscible fluids subject to attractive 
forces between them. As shown in Fig.~\ref{fig02} by the analyses of the 
incompressible and compressible parts of the kinetic energy, the vorticity dominates 
the dynamics during the first stage of time evolution, till 
$t\approx 10\omega_\perp^{-1}$, with plenty of vortex-antivortex production in the
interval $4<t<10$. For longer-time evolution, sound-wave production starts 
dominating the dynamics, with part of the kinetic energy being transferred from 
incompressible to compressible. However, the vorticity remains high enough within
the mixture (further observed in comparison to other cases we have studied). 
The dynamical overlap of the densities $\Lambda$ (shown in
Fig.~\ref{fig09}) reflects the applied immiscible condition. 
The coupled system has a small tendency to become more miscible near the time 
interval when the compressible modes start dominating the dynamics.
Our analysis in Fig.~\ref{fig03} of the corresponding spectra shows that
for both incompressible and compressible energies, the $k^{-3}$ behavior can 
be identified for $k\xi>1$, in the interval $t\approx$ 4.1 to 6.4$\omega_\perp^{-1}$.
However, in the same specific time interval,  
the classical scaling law $k^{-5/3}$ expected for turbulence on the intermediate 
$k$ region ($k\xi< 1$), can only be identified in the case of the incompressible 
kinetic energy results (dominated by the motion of quantized vortices). 
In this case, the compressible energies are not being transferred 
through a cascade process, as the incompressible ones, which is understood as
due to sound-wave radiation and dissipative effects.
At large times, $t>20\omega_\perp^{-1}$, as seen in Fig.~\ref{fig02}, the compressible and 
incompressible modes converge to constant values, with the corresponding compressible 
spectra becoming flatter in the ultraviolet limit. 
Such results imply approximate equilibrium with no significant energy transfer 
across scales.\\

\noindent 
{\bf KH dynamics:} With the same initial immiscible
condition as in the RT case ($\delta=1.05$), this instability is generated by 
applying constant forces to both components in opposite directions. 
They act parallel to the interface (borderline surface) between the species, creating a
velocity shear across the boundary. The resulting velocity differences between
the fluids trigger the KH instability, leading to characteristic rolling
patterns at the interface. This mechanism highlights the critical role of shear 
flow in driving interfacial instabilities in coupled condensates.

The main results in this case are given in Figs.~\ref{fig04}-\ref{fig06}, showing
that the KH dynamics is dominated by vorticity throughout the time evolution, 
in contrast with results obtained for the RT dynamics.
The two-component spectra over the wavenumber $k$ also present 
a behavior more uniform than the RT instability case during the onset of instability.
As shown in Fig.~\ref{fig06} for the kinetic energy spectra of both species, in 
the ultraviolet limit, the expected $k^{-3}$ behavior is noticed for both compressible
and incompressible kinetic energy cases. This behavior starts close to 
$k\xi\sim 0.5$. In the interval $k_L\xi<k\xi<0.5$, the $k^{-5/3}$ behavior is better 
observed in the compressible energy modes, deviating to $\sim k^{-2}$ in the 
incompressible cases.
Essentially, the similar scaling for both incompressible and compressible 
modes is implying that the specific external forces are equally exciting these 
modes in the KH instability case, with the constraints homogenizing the energy 
transfer processes. In the long-time evolution, the vorticity and sound-wave production 
remain approximately stable, kept by the constant forces.

As related to the evolution of the overlap between the densities, the KH instability
differs from the RT instability mainly during the onset of instabilities, as shown 
in Fig.~\ref{fig07}. Along all the KH dynamics, $\Lambda$ remains below 0.03 (less
than 3\% miscible).\\

\noindent{\bf IMQT dynamics:} Here, the critical role of nonlinearity 
changes in destabilizing the system is highlighted. The study underscores the 
interplay between interaction strength and phase separation in coupled condensates.
The instability is triggered by quenching the nonlinearity of the system, through a
sudden transition from an immiscible to a miscible condition, with the miscible 
condition remaining throughout the dynamics.

Two possible initial immiscible configurations are assumed. 
The first, with $\delta = 1.02$, having the two
species occupying three distinct regions inside the trap, in a kind of 
projected ``tennis-ball" configuration, as seen in the $t=0$ panels 
(a$_i$) of Fig.~\ref{fig08}. 
Symmetrically, the $^{85}$Rb is placed in the central part, 
with $^{87}$Rb split in the other two parts. 
The second, with $\delta = 1.05$. in an axially symmetric configuration, with 
each species having half of the trapping region, as seen in the $t=0$ panels 
(a$_i$) of Fig.~\ref{fig11}. 
In both cases, the quenching from an immiscible to a miscible system
is performed by a sudden change in the interspecies interaction, 
such that the onset of the dynamics is developed with $\delta =0.75$.
As observed in the first case, we have an asymmetric initial production of vortices, 
with the component located in the center ($^{85}$Rb) presenting less vorticity than 
the other component. This is an expected result, considering the sudden change from 
immiscible to miscible configuration, with the element outside the center, $^{87}$Rb, 
moving to the central part through two boundary edges between the species, while the 
element in the center, $^{87}$Rb, remains more confined by the pressure of the 
second component.
In the second case, with the initial configuration having both components symmetrically 
positioned, similar production of vortices and sound waves are noted throughout 
the temporal evolution, as seen in Fig.~\ref{fig12} for the incompressible kinetic 
energy part (upper panel), with the corresponding number of vortices (lower panel).
The respective spectra, given in Figs.~\ref{fig10} and \ref{fig13}, show similar 
behavior for the incompressible and compressible modes, with energy oscillations
between the two modes (indicating energy transfer) for both species. 
When averaging the oscillations, the behaviors of the incompressible mode 
(for $k\xi<0.5$) are close to $k^{-5/3}$ in the first case, as seen in 
Fig.~\ref{fig10}; and deviating slightly towards $k^{-2}$ in the case shown 
in Fig.~\ref{fig13}. The results also indicate the coupling between the 
compressible and incompressible modes, as both follow approximately the same behavior.\\
Characteristic of both IMQT instabilities and distinguishable from RT and KH 
instabilities, a kind of bottleneck effect is noted in the energy 
spectra~\cite{2004Connaughton,2007Lvov}. It arises due to mismatches between 
the energy transfer rate and the dissipation rate on small scales. Thus, before 
the onset of the dissipation range, an increase in energy occurs on intermediate 
scales, within the enstrophy cascade range, where the Kolmogorov scale breaks. 
It is interpreted as associated with nonlinear interactions, with energies 
temporarily accumulating before being dissipated on small scales.

In all processes examined, many dipoles and turbulent flows are observed in the 
binary mixture, in the onset of instabilities, which induce 
spontaneous occurrence of vortex dipoles followed by sound-wave (phonon) production.
 All the instability cases were confronted with the expected classical scaling law behaviors,
 by spectral analyses.
The Kolmogorov's scaling $k^{-5/3}$ behavior, expected in the kinetic 
energy interval in which the wavenumber is smaller than the inverse of the healing length ($k<1/\xi$),
is approximately confirmed in particular time intervals when the instabilities emerge.
This behavior appears to be more limited in the time interval for the RT case, which we
interpret as being due to the two-step perturbation procedure leading to strong 
compressible density fluctuations during the initial dynamics. 
In the ultraviolet region (for $k>1/\xi$), the $k^{-3}$ behavior is recognized at 
specific times of the onset of instabilities for all the cases, particularly
for the incompressible part of the energy. 
In the spectral analyses of the IMQT cases (when only the nonlinear interactions 
are suddenly modified), the expected classical Kolmogorov scaling behavior is also 
not followed in an intermediate region before the ultraviolet region, which is
associated with a kind of bottleneck effect occurring before the energy dissipation at small scales.
 
In summary, in this work, we have presented numerical simulations of three kinds
of instabilities in a binary coupled BEC mixture, obtained by using the coupled GP
formalism, initially prepared in immiscible configurations. 
The cases are understood as accessible for experimental realizations, considering 
the actual cold-atom facilities. The RT and KH instabilities are investigated by 
keeping the coupled condensates in an immiscible configuration along the dynamics,
whereas the IMQT instability is obtained by quenching the nonlinear two-body 
parameter, from immiscible to miscible configuration, considering two different 
initial conditions. 
Our main objective was to explore different kinds of engineered instabilities in a
comparative way, which can emerge when coupling two initial immiscible condensates.
Several interesting aspects of the dynamics, such as those related to vorticity 
in the long-term evolution, or associated with interference patterns in the coupled 
densities, are highlighted. However, such analyses are beyond the scope of the present 
work. Deep-focus investigations are demanding in such cases.
Regarding possible similarities between classical and quantum turbulence, 
the main outcome was derived from spectral analyses of numerical simulations 
that occur in the short time intervals at which the onset of the instabilities 
can be followed through the compressible and incompressible spectral behaviors. 
As expected, for longer times, the simulated dynamics for the different cases 
under study deviates from the classical one, reflecting the fact that we have 
zero viscosity in quantum fluids, with vortex dynamics and quantum 
interactions being primarily responsible for dissipation.

\section*{Acknowledgements}
We are grateful to Prof. Ashton S. Bradley for the suggestions, as well as
for local support obtained by one of us (RKK) during part of the realization of this work. 
For partial support during the realization of this work, we thank the following agencies:
Marsden Fund, Contract UOO1726 (RKK);
Funda\c c\~ao de Amparo \`a Pesquisa do Estado de S\~ao Paulo 
[Projs.~2024/04174-8 (SS and LT) and 2024/01533-7 (AG and LT)]; and
Conselho Nacional de Desenvolvimento Cient\'\i fico e Tecnol\'ogico  
[Procs. 303263/2025-3 (LT), and 306219/2022-0 (AG)].


\begin{thebibliography}{99}
\bibitem{1883-Reynolds}
O. Reynolds, 
XXIX. An experimental investigation of the circumstances which determine whether the 
motion of water shall be direct or sinuous, and of the law of resistance in parallel 
channels, Philos. Trans. R. Soc. {\bf 174}, 935 (1883).

\bibitem{1908-Sommerfeld}
 A. Sommerfeld, 
{Ein Beitrag zur hydrodynamischen Erkl\"arung der turbulenten
Fl\"ussigkeitsbewegung}, in: Proc. of the 4th Int. Mathematical Congress,
Rome 1908, {\bf 3}, 116 (1909).

\bibitem{2010-Eckert} M. Eckert, {The troublesome birth of hydrodynamic 
stability theory: Sommerfeld and the turbulence problem}, 
Eur. Phys. J. H {\bf 35}, 29 (2010).

\bibitem{Kolmogorov1941}
A. N. Kolmogorov, 
The local structure of turbulence in incompressible viscous fluid for very 
large Reynolds' numbers, Dokl. Akad. Nauk SSSR {\bf 30}, 301 (1941).

\bibitem{1995Frisch}
U. Frisch, Turbulence: The Legacy of A. N. Kolmogorov
(Cambridge University Press, Cambridge, 1995).

\bibitem{1986Donnelly} R. J. Donnelly and C. E. Swanson, {Quantum turbulence}, 
J. Fluid Mech. {\bf 173}, 387 (1986).

\bibitem{2008Barenghi} C. F. Barenghi, {Is the Reynolds number infinite in superfluid
turbulence?}, Physica D {\bf 237 D}, 2195 (2008).

\bibitem{2012Bradley} A. S. Bradley and B. P. Anderson, {Energy spectra of vortex 
distributions in two-dimensional quantum turbulence}, 
Phys. Rev. X {\bf 2}, 041001 (2012).

\bibitem{2015Reeves} M. T. Reeves, T. P. Billam, B. P. Anderson, and A. S. Bradley,
Identifying a superfluid Reynolds number via dynamical similarity,
Phys. Rev. Lett. {\bf 114}, 155302 (2015).

\bibitem{1991Donnelly} R. J. Donnelly, {\it Quantized Vortices in Helium II} 
(Cambridge University Press, Cambridge, England, 1991).

\bibitem{2013Jou} D. Jou and M. Sciacca, 
{Quantum Reynolds number for superfluid counterflow turbulence}, 
Bollettino di Matematica Pura e Applicata (M. S. Mongioví, M. Sciacca, and S. Triolo, eds.), 
vol. VI, pp. 95–103, Aracne editrice, 2013.

\bibitem{2018Mongiovi} M. S. Mongiov\`i, D. Jou, and M. Sciacca, 
{Non-equilibrium thermodynamics, heat transport and thermal waves in laminar and 
turbulent superfluid helium}, Phys. Rep. {\bf 726}, 1 (2018).

\bibitem{2002Vinen} W. F. Vinen and J. J. Niemela, Quantum turbulence, 
J. Low Temp. Phys. {\bf 128}, 167 (2002).

\bibitem{2011Paoletti} M. S. Paoletti and D. P. Lathrop, Quantum turbulence,
Annu. Rev. Condens. Matter Phys. {\bf 2}, 213 (2011).

\bibitem{2013Tsubota} M. Tsubota, M. Kobayashi, H. Takeuchi, 
{Quantum hydrodynamics}, Phys. Rep. {\bf 522}, 191 (2013).

\bibitem{2013Nemirovskii} S. K. Nemirovskii, {Quantum turbulence: 
Theoretical and numerical problems}, 
Phys. Rep. {\bf 524}, 85 (2013).

\bibitem{2014Kwon} W.J. Kwon, G. Moon, J.Y. Choi, S.W. Seo, Y.-i. Shin, 
Relaxation of superfluid turbulence in highly oblate Bose–Einstein condensates, 
Phys. Rev. A {\bf 90}, 063627 (2014).
\bibitem{2016Barenghi} C. F. Barenghi and N. G. Parker,
{A Primer on Quantum Fluids}, Springer International Publ., Switzerland, 2016

\bibitem{2016Tsatsos} M. C. Tsatsos, P. E. S. Tavares, A. Cidrim, A. R. Fritsch, 
M. A. Caracanhas, F. E. A. dos Santos, C. F. Barenghi, and V. S. Bagnato, 
{Quantum turbulence in trapped atomic Bose–Einstein condensates}, Phys. Rep.
{\bf 622}, 1 (2016).

\bibitem{2023Barenghi} C. F. Barenghi, L. Skrbek, and K. R. Sreenivasan,
Quantum Turbulence (Cambridge University Press, Cambridge, 2023).

\bibitem{2025Zhao} M. Zhao, J. Tao, and I. B. Spielman,
Kolmogorov Scaling in Turbulent 2D Bose-Einstein Condensates, 
Phys. Rev. Lett. {\bf 134}, 083402 (2025).

\bibitem{2025Fischer} T. Z. Fischer and A. S. Bradley,
Regimes of steady-state turbulence in a quantum fluid,
Phys. Rev. A {\bf 111}, 023308 (2025).

\bibitem{2025Simjanovski}
S. Simjanovski, G. Gauthier, H. Rubinsztein-Dunlop, M. T. Reeves, and T. W. Neely,
Shear-induced decaying turbulence in Bose-Einstein condensates,
Phys. Rev. A {\bf 111}, 023314 (2025).

\bibitem{2017Tsubota} M. Tsubota, K. Fujimoto, S. Yui, 
{Numerical studies of quantum turbulence}, 
J. Low Temp. Phys. {\bf 188}, 119 (2017).

\bibitem{2021Kobayashi}M. Kobayashi, P. Parnaudeau, F. Luddens, C. Lothod\'e, 
L. Danaila, M. Brachet, and I. Danaila, { Quantum turbulence simulations using 
the Gross-Pitaevskii equation: high-performance computing and new numerical benchmarks}, 
Comput. Phys. Commun. {\bf 258}, 107579 (2021). 

{
\bibitem{2017Serafini} S. Serafini, L. Galantucci, E. Iseni, T. Bienaimé, 
R. N. Bisset, C. F. Barenghi, F. Dalfovo, G. Lamporesi, and G. Ferrari, 
Vortex reconnections and rebounds in trapped Atomic 
Bose-Einstein Condensates, Phys. Rev. X {\bf 7}, 021031 (2017).}

\bibitem{Vinen1957} W. F. Vinen, 
Mutual friction in a heat current in liquid helium II 
I. Experiments on steady heat currents, Proc. R. Soc. {\bf 240}, 114 (1957); 
Mutual friction in a heat current in liquid helium II. II. Experiments on transient 
effects, Proc. R. Soc. {\bf 240}, 128 (1957);
Mutual friction in a heat current in liquid helium II III. 
Theory of the mutual friction {\bf 242}, 493 (1957).

\bibitem{Bagnato2009}
E. A. L. Henn, J. A. Seman, G. Roati, K. M. F. Magalh\~aes, and V. S. Bagnato, 
 Emergence of turbulence in an oscillating Bose-Einstein condensate,
 Phys. Rev. Lett. {\bf 103}, 045301 (2009). 

\bibitem{2014White} A. C. White, B. P. Anderson, and V. S. Bagnato,
Vortices and turbulence in trapped atomic condensates, PNAS {\bf 111}, 4719 (2014).

\bibitem{Navon2016} 
N. Navon, A. L. Gaunt, R. P. Smith, and Z. Hadzibabic, 
Emergence of a turbulent cascade in a quantum gas,
Nature {\bf 539}, 72 (2016). 

{
\bibitem{Navon2019} 
N. Navon, C. Eigen, J. Zhang, R. Lopes, A. L. Gaunt, K.
Fujimoto, M. Tsubota, R. P. Smith, and Z. Hadzibabic,
Synthetic dissipation and cascade fluxes in a turbulent
quantum gas, Science {\bf 366}, 382 (2019).

\bibitem{2024Karailiev} 
A. Karailiev, M. Gazo, M. Gałka, C. Eigen, T. Satoor, and Z. Hadzibabic, 
Observation of an inverse turbulent-wave cascade in a driven quantum gas,
Phys. Rev. Lett. {\bf 133}, 243402 (2024).}

\bibitem{2022Reeves} M. T. Reeves, K. Goddard-Lee, G. Gauthier, O. R. Stockdale,
H. Salman, T. Edmonds, X. Yu, A. S. Bradley, M. Baker, H. Rubinsztein-Dunlop,
M. J. Davis, and T. W. Neely, Turbulent relaxation to equilibrium in a two-dimensional 
quantum vortex gas, Phys. Rev. X {\bf 12}, 011031 (2022).

\bibitem{Gauthier2019} 
G. Gauthier, M. T. Reeves, X. Yu, A. S. Bradley, M. Baker,  T. A. Bell, H. R. Dunlop, 
M. J. Davis, and T. W. Neely, Giant vortex clusters in a two-dimensional 
quantum fluid, Science {\bf 364}, 1264 (2019).
 
\bibitem{Johnstone2019}
S. P. Johnstone, A. J. Groszek, P. T. Starkey, C. J. Billington, T. P. Simula, 
and K. Helmerson, 
Evolution of large-scale flow from turbulence in a two-dimensional superfluid,
Science {\bf 364}, 1267 (2019). 

\bibitem{1997Nore} 
C. Nore, M. Abid, and M. E. Brachet,
Kolmogorov turbulence in low-temperature superflows,
 Phys. Rev. Lett. {\bf 78}, 3896 (1997).

\bibitem{2005Kobayashi} M. Kobayashi and M. Tsubota, 
Kolmogorov spectrum of superfluid turbulence: Numerical analysis of the 
Gross-Pitaevskii equation with a small-scale dissipation,
Phys. Rev. Lett. {\bf 94}, 065302 (2005).

\bibitem{2009Yepez} J. Yepez, G. Vahala, L. Vahala, and M. Soe, 
Superfluid turbulence from quantum Kelvin wave to classical Kolmogorov cascades
Phys. Rev. Lett. {\bf 103}, 084501 (2009).

\bibitem{2011Baggaley} A. W. Baggaley and C. F. Barenghi, 
Spectrum of turbulent Kelvin-waves cascade in superfluid helium,
Phys. Rev. B {\bf 83}, 134509 (2011).

\bibitem{2013Reeves} M. T. Reeves, T. P. Billam, B. P. Anderson and A. S. Bradley, 
{Inverse energy cascade in forced two-dimensional quantum turbulence},
Phys. Rev. Lett. {\bf 110}, 104501 (2013).

\bibitem{2014Reeves} M. T. Reeves, T. P. Billam, B. P. Anderson and A. S. Bradley, 
Signatures of coherent vortex structures in a disordered two-dimensional quantum fluid,
Phys. Rev. A {\bf 89}, 053631 (2014).

\bibitem{1998Hall} D. S. Hall, M. R. Matthews, J. R. Ensher, C. E. Wieman, and E. A. Cornell,
{Dynamics of Component Separation in a Binary Mixture of Bose-Einstein Condensates},
Phys. Rev. Lett. {\bf 81}, 1539 (1998).

\bibitem{2002Pethick} C. J. Pethick and H. Smith, 
{Bose-Einstein Condensation in Dilute Gases} (Cambridge University Press, 
Cambridge, 2002).

\bibitem{2017kumar} R. K. Kumar,  P. Muruganandam, L. Tomio, and A. Gammal, 
Miscibility in coupled dipolar and non-dipolar Bose-Einstein condensates,
J. Phys. Commun. {\bf 1}, 035012 (2017). 

\bibitem{1900Rayleigh} L. Rayleigh, Scientific Papers, Vol. II (Cambridge Univ. Press, 
Cambridge, England, 1900), 200.

\bibitem{1950Taylor} G. I. Taylor, {The instability of liquid surfaces when 
accelerated in a direction perpendicular to their planes}, Proc. R. Soc. 
London Ser. A {\bf 201}, 192 (1950).

\bibitem{1871Thomson} W. Thompson (Lord Kelvin), Hydrokinetic solutions and observations,
Philos. Mag. {\bf 42} 362 (1871); {\bf 10}, 155 (1980).
\bibitem{1868Helmholtz} H. von Helmholtz, \"Uber discontinuierliche 
Flüssigkeits-Bewegungen, Monatsberichte der Königlichen Preussische Akademie der
Wissenschaften zu Berlin {\bf 23}, 215 (1868).

\bibitem{1984sharp} D. H. Sharp,  
An overview of Rayleigh-Taylor instability, Physica D {\bf 12}, 3 (1984).

\bibitem{2020Banerjee} A. Banerjee, {Rayleigh-Taylor instability: A status review of 
experimental designs and measurement diagnostics}, 
J. Fluids Eng. {\bf 142}, 120801 (2020).

\bibitem{2002Blaauwgeers}
R. Blaauwgeers, V. B. Eltsov, G. Eska, A. P. Finne, R. P. Haley, M. Krusius, J. J. Ruohio, 
L. Skrbek, and G. E. Volovik, {Shear flow and Kelvin-Helmholtz instability in superfluids},
Phys. Rev. Lett. {\bf 89}, 155301 (2002).

\bibitem{2006Finne} A. P. Finne, V. B. Eltsov, R. Hänninen, N. B. Kopnin, J. Kopu, 
M. Krusius, M. Tsubota and G. E. Volovik, {Dynamics of vortices and interfaces in 
superfluid $^3$He}, Rep. Prog. Phys. {\bf 69}, 3157 (2006).

\bibitem{2006Berloff} N. G. Berloff and C. Yin, Turbulence and coherent structures 
in two-component Bose condensates, J. Low Temp. Phys. {\bf 145}, 187 (2006).

\bibitem{Sasaki2009} 
K. Sasaki, N. Suzuki, D. Akamatsu, and H. Saito, 
Rayleigh-Taylor instability and mushroom-pattern formation in a two-component 
Bose-Einstein condensate,
Phys. Rev. A {\bf 80}, 063611 (2009).

\bibitem{Takeuchi2010} H. Takeuchi, N. Suzuki, K. Kasamatsu, H. Saito, and M. Tsubota, 
Quantum Kelvin-Helmholtz instability in phase-separated two-component Bose-Einstein 
condensates, Phys. Rev. B {\bf 81}, 094517 (2010).

{
\bibitem{2010Gautam} S. Gautam and D. Angom, Rayleigh-Taylor instability in 
binary condensates, Phys. Rev. A {\bf 81}, 053616 (2010).}

\bibitem{Bezett2010}
A. Bezett, V. Bychkov, E. Lundh, D. Kobyakov, and M. Marklund, 
Magnetic Richtmyer-Meshkov instability in a two-component Bose-Einstein condensate,
Phys. Rev. A {\bf 82}, 043608 (2010). 

\bibitem{Kobyakov2011}
D. Kobyakov, V. Bychkov, E. Lundh, A. Bezett, V. Akkerman,  and M. Marklund, 
Interface dynamics of a two-component Bose-Einstein condensate driven by an external force,
Phys. Rev. A {\bf 83}, 043623 (2011). 

\bibitem{Kobyakov2014} 
D. Kobyakov, A. Bezett, E. Lundh, M. Marklund, and V. Bychkov, 
Turbulence in binary Bose-Einstein condensates generated by highly nonlinear Rayleigh-Taylor 
and Kelvin-Helmholtz instabilities,
Phys. Rev. A {\bf 89}, 013631 (2014). 

\bibitem{2016Fujimoto}
K. Fujimoto and M. Tsubota, Direct and inverse cascades of spin-wave turbulence in spin-1 
ferromagnetic spinor Bose-Einstein condensates, Phys. Rev. A {\bf 93}, 033620 (2016).

\bibitem{2021Mithun}
T. Mithun, K. Kasamatsu, B. Dey, and P. G. Kevrekidis, Decay of two-dimensional quantum 
turbulence in binary Bose-Einstein condensates, Phys. Rev. A {\bf 103}, 023301 (2021).

\bibitem{2021Kokubo} 
H. Kokubo, K. Kasamatsu, and H. Takeuchi, 
Pattern formation of quantum Kelvin-Helmholtz instability in binary
superfluids, Phys. Rev. A {\bf 104}, 023312 (2021).

\bibitem{2023Saboo} 
A. Saboo, S. Halder, S. Das, and S. Majumder,  
Rayleigh-Taylor instability in a phase-separated three-component
Bose-Einstein condensate, Phys. Rev. A {\bf 108}, 013320 (2023).

\bibitem{2023Silva} A. N. da Silva, R. K. Kumar,  A. S. Bradley, 
and L. Tomio, {Vortex generation in stirred binary Bose-Einstein condensates}, 
Phys. Rev. A {\bf 107}, 033314 (2023).

\bibitem{2024Kadokura} T. Kadokura and H. Saito, 
Kolmogorov-Hinze scales in turbulent superfluids, 
Phys. Rev. Lett. {\bf 133}, 256001 (2024).

{
\bibitem{2024Zhou} Y. Zhou, Hydrodynamic instabilities and turbulence: Rayleigh-Taylor, 
Richtmyer-Meshkov, and Kelvin-Helmholtz mixing (Cambridge University Press, Cambridge, 
2024).}

\bibitem{2024Bigagli} N. Bigagli, W. Yuan, S. Zhang, 
B. Bulatovic, T. Karman, I. Stevenson, and S. Will,
Observation of Bose-Einstein condensation of dipolar molecules,
Nature {\bf 631}, 289 (2024). 

\bibitem{Sabari2024}S. Sabari, R. K. Kumar, L. Tomio, { Vortex dynamics and turbulence in 
dipolar Bose-Einstein condensates}, 
Phys. Rev. A {\bf 109}, 023313 (2024).

\bibitem{2018Sabari} S. Sabari and R. K. Kumar, 
{Effect of an oscillating Gaussian obstacle in a dipolar Bose-Einstein condensate}, 
Eur. Phys. J. D \textbf{72}, 48 (2018).

\bibitem{Lauro2024}L. Tomio, A. N. da Silva, S. Sabari, R. K. Kumar,  
{Dynamical vortex production and quantum turbulence in perturbed Bose-Einstein
condensates}, Few-Body Systems {\bf 65}, 13 (2024).

\bibitem{2001Kivotides} D. Kivotides, C. F. Barenghi, and D. C. Samuels, 
Fractal dimension of superfluid turbulence,
Phys. Rev. Lett. {\bf 87}, 155301 (2001).

\bibitem{2020Koniakhin} S. V. Koniakhin, O. Bleu, G. Malpuech, and D. D. Solnyshkov,
2D quantum turbulence in a polariton quantum fluid,
Chaos, Solitons and Fractals {\bf 132}, 109574 (2019).

\bibitem{1985Constantin}  P. Constantin, C. Foias, O. P. Manley and R. Temam, 
Determining modes and fractal dimension of turbulent flows,
J. Fluid Mech. {\bf 150}, 427 (1985).

\bibitem{2009Rakhshandehroo} G. R. Rakhshandehroo, M. R. Shaghaghian,  
A. R. Keshavarzi, N. Talebbeydokhti,
Temporal variation of velocity components in a turbulent open channel flow: Identification 
of fractal dimensions, Appl. Math. Model. {\bf 33}, 3815 (2009).

 \bibitem{2024Geng} Y. Geng, J. Tao, S. Mukherjee, S. Eckel, G. K. Campbell, and
 I. B. Spielman, {The Rayleigh-Taylor instability in a binary quantum fluid},
 arXiv: 2411.19807v1. 

\bibitem{2024Bayocboc} F. A. Bayocboc Jr., J. Dziarmaga, and W. H. Zurek,
Biased dynamics of the miscible-immiscible quantum phase transition in a binary 
Bose-Einstein condensate, Phys. Rev. B {\bf 109}, 064501 (2024).

\bibitem{Mukherjee2020} K. Mukherjee, S. I. Mistakidis, P. G. Kevrekidis, and P. Schmelcher, 
Quench induced vortex-bright-soliton formation in binary Bose-Einstein 
condensates, J. Phys. B: At. Mol. Opt. Phys. {\bf 53}, 055302 (2020).

\bibitem{2016Eto} 
Y. Eto, M. Takahashi, M. Kunimi, H. Saito, and T. Hirano,
Nonequilibrium dynamics induced by miscible–immiscible transition in binary Bose–Einstein 
condensates, New J. Phys. {\bf 18}, 073029 (2016).

\bibitem{2004Kasamatsu} K. Kasamatsu and M. Tsubota,
Multiple Domain Formation Induced by Modulation Instability in Two-Component 
Bose-Einstein Condensates, Phys. Rev. Lett. {\bf 93}, 100402 (2004).

\bibitem{2019Thiruvalluvar}R.T. Thiruvalluvar, E. Wamba, S. Sabari, K. Porsezian, 
{Impact of higher-order nonlinearity on modulational instability in two-component 
Bose-Einstein condensates}, Phys. Rev. E {\bf 99}, 032202 (2019).

\bibitem{1955Feynman} R. P. Feynman, Application of Quantum Mechanics to
Liquid Helium, Progress in Low Temperature Physics,
Vol. I, edited by C. J. Gorter (North-Holland, Amsterdam, 1955).
 
\bibitem{2022Bradley} A. S. Bradley, R. K. Kumar, S. Pal, and X. Yu,  
{Spectral analysis for compressible quantum fluids}, 
Phys. Rev. A {\bf 106}, 043322 (2022).

\bibitem{1999Timmermans} E. Timmermans, P. Tommasini, M. Hussein, and A. Kerman, 
{Feshbach resonances in atomic Bose-Einstein condensates}, Phys. Rep. {\bf 315}, 199
(1999).

\bibitem{2010Chin} C. Chin, R. Grimm, P. Julienne, and E. Tiesinga, 
Feshbach resonances in ultracold gases, Rev. Mod. Phys. {\bf 82}, 1225 (2010).

\bibitem{2019Kumar} R. K. Kumar,  L. Tomio, and A. Gammal, 
Vortex patterns in rotating dipolar Bose–Einstein condensate mixtures 
with squared optical lattices, 
J. Phys. B:  At. Mol. Opt. Phys. {\bf 52}, 025302 (2019). 

\bibitem{Kishor2020} R. K. Kumar, A. Gammal, and L.  Tomio, 
Mass-imbalanced Bose-Einstein condensed mixtures in rotating perturbed trap,
Phys. Lett. A {\bf 384}, 126535 (2020). 

\bibitem{2022Mukherjee} B. Mukherjee, A. Shaffer, P. B. Patel, Z. Yan, C. C. Wilson,
V. Crépel, R. J. Fletcher, and M. Zwierlein, 
Crystallization of bosonic quantum Hall states in a rotating quantum gas,
Nature {\bf 601}, 58 (2022).

\bibitem{2024Rajkov} D. D. Hernández-Rajkov, N. Grani, F. Scazza, G. Del Pace, 
W. J. Kwon, M. Inguscio, K. Xhani, C. Fort, M. Modugno, F. Marino, and G. Roati, 
Connecting shear flow and vortex array instabilities in annular atomic superfluids, 
Nature Physics {\bf 20}, 939 (2024).

\bibitem{2024Huh} S. Huh, W. Yun, G. Yun, S. Hwang, K. Kwon, J. Hur, S. Lee, H. Takeuchi,
S. K. Kim, and J.-Y. Choi, Beyond skyrmion spin texture from quantum Kelvin-Helmholtz
instability, arXiv: 2408.11217v1.

{
\bibitem{2011Aluie} H. Aluie, Compressible turbulence: The cascade and its locality,
Phys. Rev. Lett. {\bf 106}, 174502(2011).

\bibitem{Suppl} See Supplemental Material at http://link.aps.org/
supplemental/ with animations related to 
Figs.\ref{fig01}, \ref{fig04}, \ref{fig08}, and \ref{fig11}.
}

\bibitem{2011McCarron} D. J. McCarron, H. W. Cho, D. L. Jenkin, M. P. K\"oppinger and 
S. L. Cornish, Dual-species Bose-Einstein condensate of $^{87}$Rb and  $^{133}$Cs, 
 Phys. Rev. A {\bf 84}, 011603(R) (2011).

\bibitem{2005Kozik} E. Kozik and B. Svistunov,
Vortex-phonon interaction, Phys. Rev. B {\bf 72}, 172505 (2005).

\bibitem{2016Mendonca} J. T. Mendonça, F. Haas and A. Gammal, 
Nonlinear vortex-phonon interactions in a Bose–Einstein condensate,
J. Phys. B: At. Mol. Opt. Phys. {\bf 49}, 145302 (2016).

\bibitem{2018Bradley} A. S. Bradley, Dunedin, New Zealand, 2018,
{\small //github.com/AshtonSBradley/VortexDistributions.jl.}

\bibitem{1997Andrews} M. R. Andrews, C. G. Townsend, H.-J. Miesner, D. S. Durfee, 
D. M. Kurn, and W. Ketterle, Observation of interference between two Bose condensates,
Science {\bf 275}, 637 (1997).

\bibitem{2010Frantzeskakis} D. J. Frantzeskakis, Dark solitons in atomic 
Bose–Einstein condensates: from theory to experiments, J. Phys. A: Math. Theor. 
{\bf 43}, 213001 (2010).

\bibitem{1995Pelinovsky} D. E. Pelinovsky, Y. A. Stepanyants, and Y. S. Kivshar,
Self-focusing of plane dark solitons in nonlinear defocusing media, Phys. Rev. 
E {\bf 51}, 5016 (1995).

\bibitem{2006El} G. A. El, A. Gammal, and A. M. Kamchatnov,
Oblique Dark Solitons in Supersonic Flow of a Bose-Einstein Condensate,
Phys. Rev. Lett. {\bf 97}, 180405 (2006).

\bibitem{2012Yan} D. Yan, J. J. Chang, C. Hamner, M. Hoefer, P. G. Kevrekidis, 
P. Engels, V. Achilleos, D. J. Frantzeskakis, and J. Cuevas,
Beating dark–dark solitons in Bose–Einstein condensates,
J. Phys. B: At. Mol. Opt. Phys. {\bf 45} 115301 (2012).

\bibitem{1996Tikhonenko} V. Tikhonenko, J. Christou, B. Luther-Davies, and Y. S. Kivshar,
Observation of vortex solitons created by the instability of
dark soliton stripes,
Opt. Lett. {\bf 21},  1129 (1996).

{
\bibitem{2013Cetoli} A. Cetoli, J. Brand, R. G. Scott, F. Dalfovo, and L. P. Pitaevskii,
Snake instability of dark solitons in fermionic superfluids, Phys. Rev. A
{\bf 88}, 043639 (2013).

\bibitem{2010Wen} L. Wen, H. Xiong, B. Wu, 
Hidden vortices in a Bose-Einstein condensate in a rotating double-well 
potential, Phys. Rev. A {\bf 82}, 053627 (2010).

\bibitem{2017Sabari} S. Sabari, Vortex formation and hidden vortices in dipolar
Bose–Einstein condensates, Phys. Lett. A {\bf 381} 3062 (2017).

\bibitem{2024Caldara} M. Caldara, A. Richaud, M. Capone, P. Massignan,
Suppression of the superfluid Kelvin-Helmholtz instability due to massive vortex cores, 
friction and confinement, SciPost Phys. {\bf 17}, 076 (2024).
}

\bibitem{2004Connaughton} C. Connaughton and S. V. Nazarenko,
Warm cascades and Anomalous Scaling in a Diffusion Model of Turbulence,
Phys. Rev. Lett. {\bf 92}, 044501 (2004).

\bibitem{2007Lvov} V. S. L'vov, S. V. Nazarenko, and O. Rudenko,
Bottleneck crossover between classical and quantum superfluid turbulence,
Phys. Rev. B {\bf 76}, 024520 (2007).
 
{
\bibitem{2009Bos} W.J.T. Bos
and J.P. Bertoglio,
Large-scale bottleneck effect in two-dimensional turbulence, 
Journal of Turbulence (JoT) {\bf 10}, 1 (2009).
}
\bibitem{2024Lima} A. del Río-Lima, J. A. Seman, R. Jáuregui, and F. J. Poveda-Cuevas,
Spatial and temporal periodic density patterns in driven Bose-Einstein condensates,
Phys. Rev. A {\bf 110}, 053318 (2024).

\end{thebibliography}
\end{document}